\documentclass[iop,twocolappendix]{emulateapj}

\newcommand{\lya}{Ly$\alpha$}

\usepackage{color}

\shorttitle{LAEs at z $\sim$ 2.8: 1. Large Scale Structure}
\shortauthors{Zhen-Ya Zheng,  et al. 2015}

\begin{document}


\title{ L\lowercase{y}$\alpha$ Emitter Galaxies at \lowercase{$z$}$\sim$ 2.8 in the Extended {\it Chandra} Deep Field-South: I. Tracing the Large-Scale Structure via L\lowercase{y}$\alpha$ Imaging$^{*}$
}


\author{Zhen-Ya Zheng\altaffilmark{1,2,3$^{a,c}$}}
\author{Sangeeta Malhotra\altaffilmark{2} }
\author{James E. Rhoads\altaffilmark{2}}
\author{Steven L. Finkelstein\altaffilmark{4}}
\author{Jun-Xian Wang\altaffilmark{5}}
\author{Chun-Yan Jiang\altaffilmark{6,7,3}}
\author{Zheng Cai\altaffilmark{8$^b$}}

\affil{$^1$Instituto de Astrofisica, Pontificia Universidad Catolica de Chile, 7820436 Santiago, Chile;  {zzheng@astro.puc.cl} \\
$^2$School of Earth and Space Exploration, Arizona State University, Tempe, AZ 85287; {sangeeta.malhotra@asu.edu}, {james.rhoads@asu.edu}\\
$^3$ Chinese Academy of Sciences South America Center for Astronomy, 7591245 Santiago, Chile \\
$^4$Department of Astronomy, The University of Texas, Austin, TX 78712; {stevenf@astro.as.utexas.edu}\\
$^5$CAS Key laboratory for Research in Galaxies and Cosmology, Department of Astronomy, \\University of Science and Technology of China, Hefei, Anhui 230026, China; {jxw@ustc.edu.cn}\\
$^6$CAS Key Laboratory for Research in Galaxies and Cosmology, Shanghai Astronomical Observatory, \\Nandan Road 80, Shanghai 200030, China; {cyjiang@shao.ac.cn}\\
$^7$Nucleo de Astronomia de la Facultad de Ingenieria, Universidad Diego Portales, Av. Ejercito Libertador 441, Santiago, Chile \\
$^8$UCO/Lick Observatory, University of California, 1156 High Street, Santa Cruz, CA 95064, USA {zcai@ucolick.org}}


\altaffiltext{*}{This paper includes data gathered with the 6.5 meter Magellan Telescopes located at Las Campanas Observatory, Chile.}
\altaffiltext{a}{Visiting astronomer, Cerro Tololo Inter-American Observatory, National Optical Astronomy Observatory, which is operated 
by the Association of Universities for Research in Astronomy (AURA) under a cooperative agreement with the National Science Foundation.}
\altaffiltext{b}{Hubble Fellow.}
\altaffiltext{c}{Present address:  Instituto de Astrofisica, 
Pontificia Universidad Catolica de Chile,  Vicuna Mackenna 4860, 7820436 Macul, Santiago, Chile}


\begin{abstract}
We present a narrowband survey with three adjacent filters for $z=$\,2.8--2.9 Lyman Alpha Emitter (LAE) 
galaxies in the Extended Chandra Deep Field South (ECDFS), along with spectroscopic followup. 
With a complete sample of 96 LAE candidates in the narrowband NB466, 
we confirm a large-scale structure at $z\sim$ 2.8 hinted from previous spectroscopic surveys. 
Compared to the blank field detected with the other two narrowband filters NB470 and NB475, the LAE density 
excess in NB466 (900 arcmin$^2$) is $\sim$ 6.0\,$\pm$\,0.8 times the standard deviation expected at $z\sim 2.8$, 
assuming a linear bias of 2. The overdense large scale structure in NB466 can be decomposed into 4 protoclusters, 
whose overdensities (each within an equivalent comoving volume $15^3$ Mpc$^3$) relative to the blank field 
(NB470+NB475) are in the range of 4.6\,--\,6.6. These 4 protoclusters are expected to evolve into a Coma-like cluster ($M \geq$ 10$^{15}$ 
M$_\odot$) at $z\sim$ 0. In the meanwhile, we investigate the various properties of LAEs at $z=$\,2.8--2.9 
and their dependence on the environment. The average star-formation rates derived from Ly$\alpha$, rest-frame 
UV and X-ray are $\sim$4, 10, and $<$16 M$_\odot$/yr, respectively, implying a \lya\ escape fraction of 
25\% $\lesssim$ $f^{Ly\alpha}_{\rm ESC}$ $\lesssim$ 40\% and  a UV continuum escape fraction 
of $f^{\rm UV,cont}_{\rm ESC}$  $\gtrsim$ 62\% for LAEs at $z$\,$\sim2.8$. The \lya\ photon density 
calculated from the integrated \lya\ luminosity function in the overdense field (NB466) is $\sim$50\% higher 
than that in the blank field (NB470+NB475), and more bright LAEs are found in the overdense field. The 3 
brightest LAEs, including a quasar at $z=$2.81, are all detected in X-ray and in NB466. These three LAE-AGNs 
contribute an extra 20--30\% \lya\ photon density compared with that from other LAE galaxies. 
Furthermore, we find that LAEs in overdense regions have larger EW values, bluer $U$\,-\,$B$ and 
$V$\,-\,$R$ ($\sim$\,2--3$\sigma$) colors compared with those in lower density regions, indicating that LAEs 
in overdense regions are younger and possible less dusty. We conclude that the structure at $z$\,$\sim$\,2.8 in the 
ECDFS field is a very significant and rare density peak similar to the SSA 22 protocluster, and narrowband imaging is 
an efficient method of detecting and studying such structures in the high-$z$ universe. 
\end{abstract}



\section{Introduction}
\label{sec:intro}

It is essential to study the galaxy evolution at $z> 2$, when a large fraction of present day 
massive galaxies are still forming stars vigorously, the activities of star formation and active galactic nuclei (AGN) increase with time, and the galaxy clusters are at the early stage of assembling galaxies, revealing the environmental effects.  
To this end, a large sample 
of high-redshift galaxies is needed. There are two main techniques to hunt for 
high-redshift star-forming galaxies, the dropout technique and the \lya-line search technique. 
The former is known as the Lyman-break technique \citep{Steidel96}, and is applied 
using deep broadband images wherein high-redshift galaxies are identified via a flux 
discontinuity caused by absorption from neutral gas in the inter-galactic medium (IGM). 
The latter method is designed to search for the strong \lya\ emission line, using deep 
narrowband images to identify galaxies where the \lya\ line is redshifted to windows of 
low night-sky emission. The \lya\ emitters (LAEs) are typically younger, lower mass, less 
dusty, and more compact than Lyman-break galaxies (LBGs). Compared to the Lyman-break 
technique, the \lya\ technique can select galaxies within a small redshift range ($\delta z/(1+z)\,\sim$\,1--2\%), 
which is more sensitive to the large-scale structure implied from the simulation work. 

 As a prominent tracer of ionizing photons,  \lya\ emission is an easy handle for the detection of both 
 star--forming galaxies  and AGN at redshifts $z$ $>$ 2. Narrowband searches 
 for high redshift \lya\ emission have been successfully conducted at a number of redshifts from 2.1 to 
 6.5 \citep[e.g., ][]{Guaita10, Ouchi08, MR02, Fink09, Hu10, Ka11, Ouchi10}. Unlike luminous AGN, 
 which inhabit massive halos in high density peaks \citep[e.g., ][]{Gilli03}, LAEs are thought to be located 
 in the lower-mass dark matter halos at redshifts $2<z<5$ \citep{Ouchi03,Gawiser07,Kovac07,Guaita10}.
 However, only more than a few high-density regions of LAEs and star-forming galaxies at high redshift have been 
 reported \citep{Keel99,Steidel00,Steidel05, Shimasaku03, Palunas04, Hayashino04, Matsuda04, Matsuda09, 
 Matsuda10, Ouchi05, Wang05, Kajisawa06, Miley06, Hatch08, Overzier08, Digby10, Kuiper10, Yang10, 
Toshikawa12, Toshikawa14, Yamada12}. Among them, the structure at $z = $3.09 in the SSA22 field is one of the most well studied regions. The large 
overdensity of SSA22 was firstly discovered in the redshift distribution of the LBGs by \citet{Steidel98}, and then
  confirmed by the spatial distribution of LAEs selected from subsequent narrowband surveys \citep{Steidel00, Hayashino04, 
  Matsuda04, Yamada12}.
 Therefore, narrowband imaging is an ideal tool to find large-scale structures and overdense regions such as protoclusters at high-redshift (i.e., $z$\,$>$\,2).

\begin{deluxetable*}{lcccccc}
\tabletypesize{\scriptsize}
\tablecaption{Properties of optical photometric data used in this work. Note that except for the GOODS-VIMOS data, which mainly cover the CDFS region, 
all other data cover the whole Extended CDFS region. \label{photo}}
\tablewidth{0pt}
\tablehead{
\colhead{ Band (filter)} & \colhead{Instrument} & \colhead{Exposure}  & \colhead{m$_{AB}$(lim)} & \colhead{$FWHM$} &  \colhead{Source} & \colhead{Ref.} \\
&  &  [ks] & (5$\sigma$)  & [arcsec] &   & 
}
\startdata
NB (NB466) 	& 4m Mosaic II	& 22.2 & 25.3 & 1.15 & This  work  & This  work  	\\ 
NB (NB470) 	&4m Mosaic II 	&	 23.4 &  25.6 & 1.09 &  --  & --   \\
NB (NB475) 	&4m Mosaic II	&	 20.7 & 25.5 & 1.06 &  -- & -- 	\\
U (U50)  	& 2.2m WFI	 &  43.6 &  26.0 & 1.07 &  EIS & Arnouts+01	\\ 
U (U$_V$)   & VLT/VIMOS & 	 94 &  27.8--28.4 & 0.8 & GOODS-VIMOS  & Nonino+09	\\ 
B (B99) 	& 2.2m WFI	 &	 69.4 & 27.3 & 0.99 &  GaBoDS & Hildebrandt+05	\\ 
V (V89)       & 2.2m WFI     &  56.0 & 27.0  & 0.93 &  GaBoDS & Hildebrandt+05 \\
R (R$_c$162)       & 2.2m WFI     &  57.1 & 27.2 & 0.81 &  GaBoDS & Hildebrandt+05 \\
R (R$_V$)     & VLT/VIMOS &  50 &  26.5--27.6   & 0.75 &  GOODS-VIMOS & Nonino+09	
\enddata
\end{deluxetable*}

\begin{figure}
\includegraphics[width=\linewidth]{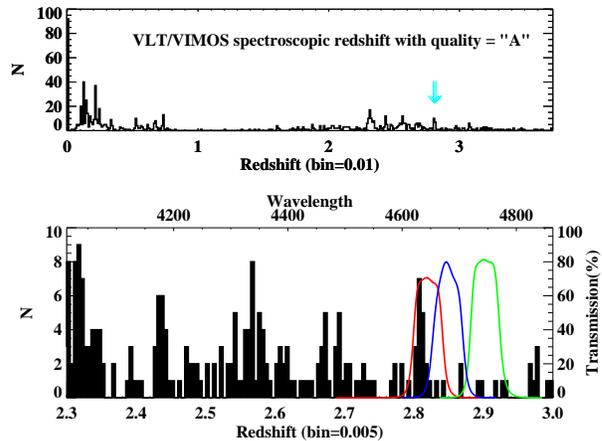} 
\caption{The VLT/VIMOS redshift distribution from Popesso et al. (2009) and Balestra et al. (2010) and the transmission 
curves of three narrowband filters (Red: NB466, Blue: NB470, and Green: NB475). Upper panel: VLT/VIMOS redshift 
distributions in the redshift range of 0 -- 4. The cyan arrow indicates the observed peak at $z\sim$ 2.8 in this distribution, 
which is covered by our NB466 narrowband imaging. Lower panel: zoom in of the distributions in the redshift range of 
2.3--3.0, and the narrowband filter transmission curves as a function of \lya\ redshift. }
\label{f1}
\end{figure}

Here we report a narrowband imaging survey and spectroscopic followup of a large-scale structure suggested 
by the redshift distributions of LBGs at $z\sim$ 2.8 in CDFS (see Fig. \ref{f1}). The redshift distributions from CDFS-VIMOS 
projects \citep{Popesso09, Balestra10} imply a large scale structure at $z\sim2.8$, 
along with one quasar at that redshift \citep{Szokoly04}. To check if it is a structure similar to SSA22, we apply 
three contiguous  narrowband filters of  NB466, NB470, and NB475, to search for LAEs at redshifts of 2.8--2.9. 
In Section 2,  we introduce our  narrow-band observations, review the techniques used to detect emission-line galaxies 
and LAEs,  and present our spectroscopic observations. We also cross-match the catalogs from {\it Chandra} X-ray 
telescope, {\it GALEX} UV telescope, and other public spectroscopic surveys in Section 2. In Section 3,  the spectroscopic 
results of LAEs confirmed at $z\sim$2.8--2.9 and their stacked spectrum are reported. We discuss the star-formation rate 
from X-ray, UV, and \lya\ for the LAEs without X-ray detections in Section 4. In Section 5 we present the Ly$\alpha$ 
luminosity function at $z\sim$2.8--2.9, and compare with other narrowband 
surveys for LAEs at $z\sim$ 2--3 . Finally, we explore the large-scale structure found via LAEs at $z\sim$ 2.8--2.9 in Section 6. 
Through out this work, we assume a cosmology with $H_0$ = 70 km s$^{-1}$ Mpc$^{-1}$, $\Omega_m$ = 0.27, and 
$\Omega_\Lambda$ = 0.73 \citep{Komatsu11}. At redshift $z=$ 2.8, the age of the Universe was  2.34 Gyr. This gives a scale of 
8.01 kpc/$\arcsec$, and a redshift change of $\delta z$ = 0.04 implies a comoving distance of 42 Mpc. 
The \citet{Salpeter55} IMF is assumed throughout the paper, and all the magnitudes are given in {\it AB} system.

\section{Data Handling}
\label{sec:data}

\subsection{Deep Narrowband Imaging}

We observed the CDF-S field with three consecutive narrowband filters with central wavelengths $\lambda_c$ of 466.6, 470.3, 
and 476.4 nm. The transmission FWHMs of the three filters are $\sim$ 50 \AA, with peak throughputs of 71.5\% for 
NB466, 80.7\% for NB470 and 82.2\% for NB475, respectively (see Fig. \ref{f1} for the transmission curves 
of the three narrowband filters). The deep narrowband images were obtained using the Mosaic \uppercase\expandafter{\romannumeral2} 
CCD imager at the Cerro Tololo Inter-American Observatory (CTIO) 4m V. M. Blanco telescope, on 2011 Oct 23-26 
(NOAO 2011B-0569, PI: Zheng).  The Blanco MOSAIC  \uppercase\expandafter{\romannumeral2}  camera has an 
array of eight 2048$\times$4096 pixel CCDs, each of which can be read out through one or two amplifiers. The CCDs 
are combined to form a 8192$\times$8192 pixel image corresponding to a field of view of 36\arcmin$\times$36\arcmin\ on 
the sky. For our observation runs, one amplifier was found to be inoperable, then we were forced to read out MOSAIC 
\uppercase\expandafter{\romannumeral2}  with 8 amplifiers only. This gererated a readout overhead of 2m40s per image. 
We obtained our deep narrowband images for each filter by combining a series of dithered 900--1200 second exposures.
We used the re-projected science frames automatically created by the NOAO Mosaic pipeline. Weights for image stacking 
were determined using weighting factor = $T$/(FWHM$^2$$\times$$\sigma_{Sky}^2$) (here $T$ is the transparency, 
FWHM is the seeing, and $\sigma_{Sky}$ is the Poisson error of sky signal), and the task $mscstack$ in the MSCRED 
package \citep{Valdes98} was used to stack the individual exposures.

The total integrated exposure times of the final stacked images are 22.2ks for NB466, 23.4ks for NB470, and 
20.7ks for NB475 in ECDFS, with average seeings of 1.15\arcsec, 1.09\arcsec, and 1.06\arcsec, respectively.  
We use MUSYC \citep{Gawiser06} $B$-band images for narrowband calibration and emission line selection ($B$-$NB$ 
color in Sec. \ref{sec:data}), and MUSYC-$U$ plus VIMOS-$U$ \citep{Nonino09} for LAE selection ($U$-$B$ color in 
Sec. \ref{sec:lyasel}). The overlap area between the narrowband stacks and the broadband images are $\sim$900 
arcmin$^2$ (32\arcmin$\times$32\arcmin\ minus 8\arcmin$\times$16\arcmin).  The redshift ranges of the LAE surveys 
with narrowband NB466, NB470, and NB475 filters are 2.800--2.842, 2.829--2.871 and 2.883--2.925 
(calculated from filter FWHM, and noting the wavelength shift of  -15\AA\ for CTIO 4m f/2.9), corresponding to LAE 
survey comoving radial scales of [44, 43, 42] Mpc,  and comoving volumes of [128950, 125678,  125349] Mpc$^3$. A 
summary of the images is given in Table \ref{photo}.

\subsection{Emission-Line Galaxy  Selection}
\label{sec:data2}

\begin{figure*}
\includegraphics[angle=0,width=\linewidth]{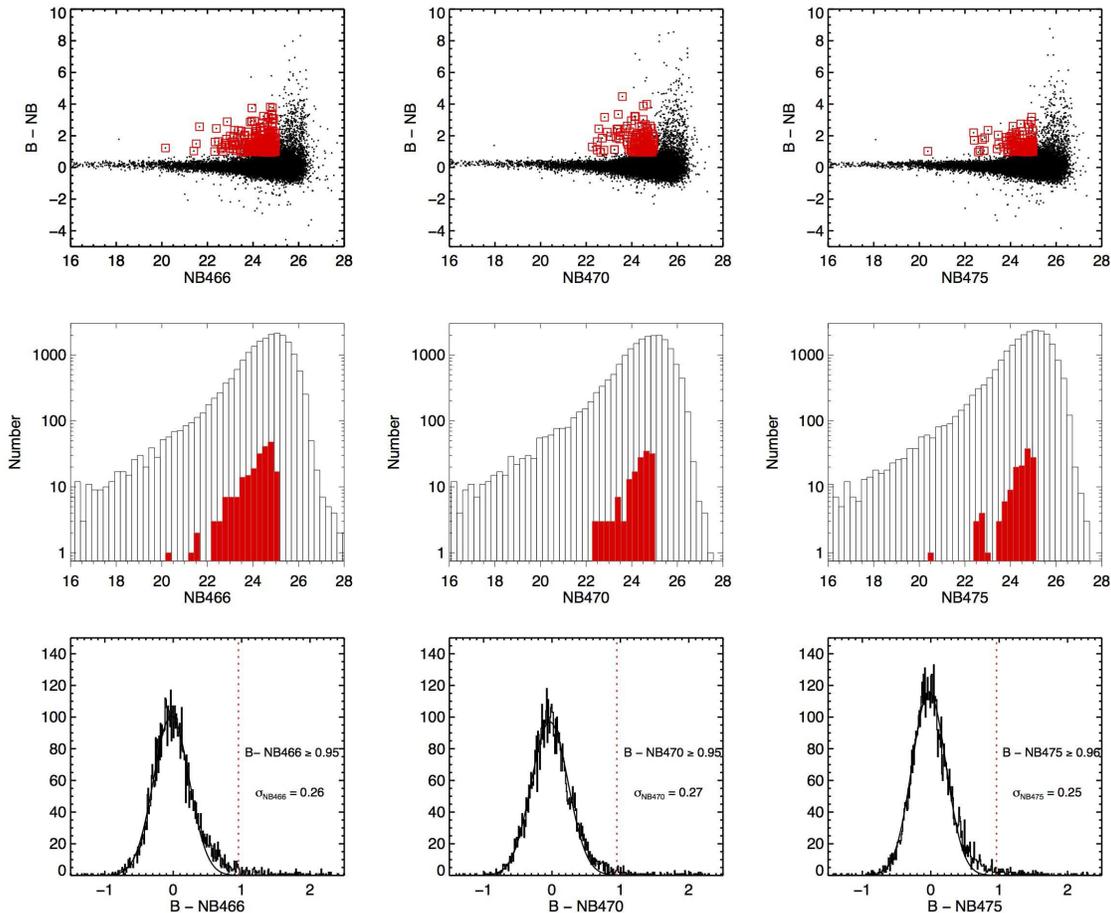} 
\caption{Top: Broad minus narrowband ($B-NB$) color vs.  narrowband magnitudes $NB$ for SExtractor-detected sources in the corresponding 
narrowband images. The black dots are narrowband detected sources, and the red squares are narrowband selected emission line galaxies with 
narrowband over B band excess of $\gtrsim$\,1\,mag, corresponding to rest-frame EW(Ly$\alpha$) $\gtrsim$\,20\,\AA\ if these candidates are 
LAEs at $z$\,$\sim$\,2.8; Middle: narrowband magnitude distributions for all detected (blank histogram) and emission line selected sources 
(red filled histogram); Bottom: Broad minus narrowband ($B-NB$) color distributions in the corresponding narrowband images. The gaussian 
fits are obtained considering the range -0.5\,$\leq$\,$B$-$NB$\,$\leq$\,0.5. The $\sigma$ of the best-fit Gaussian curves are $\sim$ 0.26. This 
means that our color cut $B$-$NB$\,$\geq$\,0.95 is selecting objects above 3.6$\sigma$.      }
\label{nbsel}
\end{figure*}

We use SExtractor \citep[version 2.8.6, ][]{Bertin96} to detect sources on the narrowband images, and SExtractor's 
two-image mode to measure the MUSYC broad band photometry of the narrowband detected sources. All fluxes are measured 
in AUTO magnitude, which yields about twice the fluxes measured from the 2\arcsec.14 (8 pixel) diameter aperture.  
The magnitude distribution of each narrowband is plotted in Figure \ref{nbsel}.  Sometimes the narrowband-selected objects are too faint 
to be detected significantly in the continuum image. In calculating and analyzing the colors or equivalent widths of these objects, 
we substitute these with their 1$\sigma$ limiting aperture magnitudes.

The emission line candidates are selected as the targets for spectroscopic followup, which should have 
(1) narrowband detection at $>$\,5$\sigma$ significance; 
(2) narrowband excess over B band $B$\,-\,$NB$\,$\geq$\,0.95 mag, so that EW$_{obs}$\,$\geq$\,76\AA\ 
(corresponding to EW$_{Ly\alpha, rest}$\, $\geq$\,20\AA\ for $Ly\alpha$ at $z$\,$=$\,2.8, see Appendix for details); 
and (3) significance of narrowband excess $>$\,4$\sigma$. 
These candidates include LAEs at $z$\,$\sim$\,2.8--2.9,  interlopers such as [O\,{\sc ii}] emitter galaxies at $z$\,$\sim$\,0.25, AGN with [C\,{\sc iv}] at 
$z$\,$\sim$\,2.0, and AGN with [Mg\,{\sc ii}] at $z$\,$\sim$\,0.67. Because the density of [O\,{\sc ii}] emitter galaxies is much higher than AGN, the main 
interlopers should be  [O\,{\sc ii}] emitter galaxies. We estimate the number of [O\,{\sc ii}] emitters from the complete sample of emission-line galaxies 
from the Hubble Space Telescope Probing Evolution and Reionization Spectroscopically Grism Survey \citep[PEARS,][]{Pirzkal13}. PEARS had found 
269 [O\,{\sc ii}] emitters in the redshift range of 0.5 $<z<$ 1.6 in an area of 119\,arcmin$^2$, 40\% of which have EW([O\,{\sc ii}]) $\geq$ 60\AA\ ($\sim$76\AA/(1+0.25)). 
Assuming that the number density of emitters did not evolve with redshift in the redshift range of 0.5 $<z<$ 1.6, we estimate $\sim$\,1--5 [O\,{\sc ii}] 
emitters in our narrowband sample\footnote{The number 1 and 5 are calculated from the subsample with both F[O\,{\sc ii}]$\geq F_{lim}$ and EW([O\,{\sc ii}]) 
$\geq$ EW$_{lim}$, and the subsample with only EW([O\,{\sc ii}]) $\geq$ EW$_{lim}$. }. The Lyman-break cut for LBGs at $z$\,$\sim$\,3 is $U$-$B$\,$>$\,1 
\citep[e.g.,][]{Hildebrandt05}. However, for galaxies at $z\sim2.8$, the $U$-$B$ cut decreases, and it can be as low as 0.3 (0.4) with VIMOS-$U$ (MUSYC-$U$ or WFI-$U$). 
We would like to check the fraction of interlopers and the color cut for LAE selection at $z$\,$\sim$\,2.8 with the spectroscopic results. Therefore we put 30\% of 
the emission line candidates selected randomly on the multi-slit masks, and test the completeness fraction and success fraction for LAE selection at $z$\,$\sim$\,2.8 
with Magellan/IMACS spectroscopic followup (see section 2.3 and 2.4). The depths (5\,$\sigma$ limiting magnitude, see Tab. \ref{photo}) of our narrow 
 bands are in the range of 25.3--25.6. We require a global narrowband limit of NB $\leq$ 25.0 for all the LAE candidates, corresponding to \lya\ flux f($Ly\alpha$) $\geq$ 
 2.9$\times$10$^{-17}$ erg s$^{-1}$ cm$^{-2}$, and luminosity L($Ly\alpha$) $\geq$ 42.3 for B-band non-detection.

With the selection criteria above and visual inspection individually, we select 217, 147, and 134 emission-line candidates in the NB466, NB470, and NB475 band, respectively.  
There are 27 candidates selected in two narrow bands, of which 16 in both NB466 and NB470,  7 in both NB470 and NB475, and 4 in both NB466 and NB475. There are 
also 5 candidates selected in all three narrow bands, and one of them is already confirmed as a Mg\,{\sc ii} at $z$\,=\,0.68 with X-ray detection.  We assign these multi-band selected 
candidates to their narrowband with maximum signals.  After excluding the duplicates, there are 461 emission-line candidates in total (204, 134, and 123 in the NB466, NB470, and NB475 band, respectively). The LAE candidates are selected from these emission-line candidates with an extra color criterion introduced in Sec. \ref{sec:lyasel}.

 \begin{figure*}
\includegraphics[angle=270,width=\linewidth]{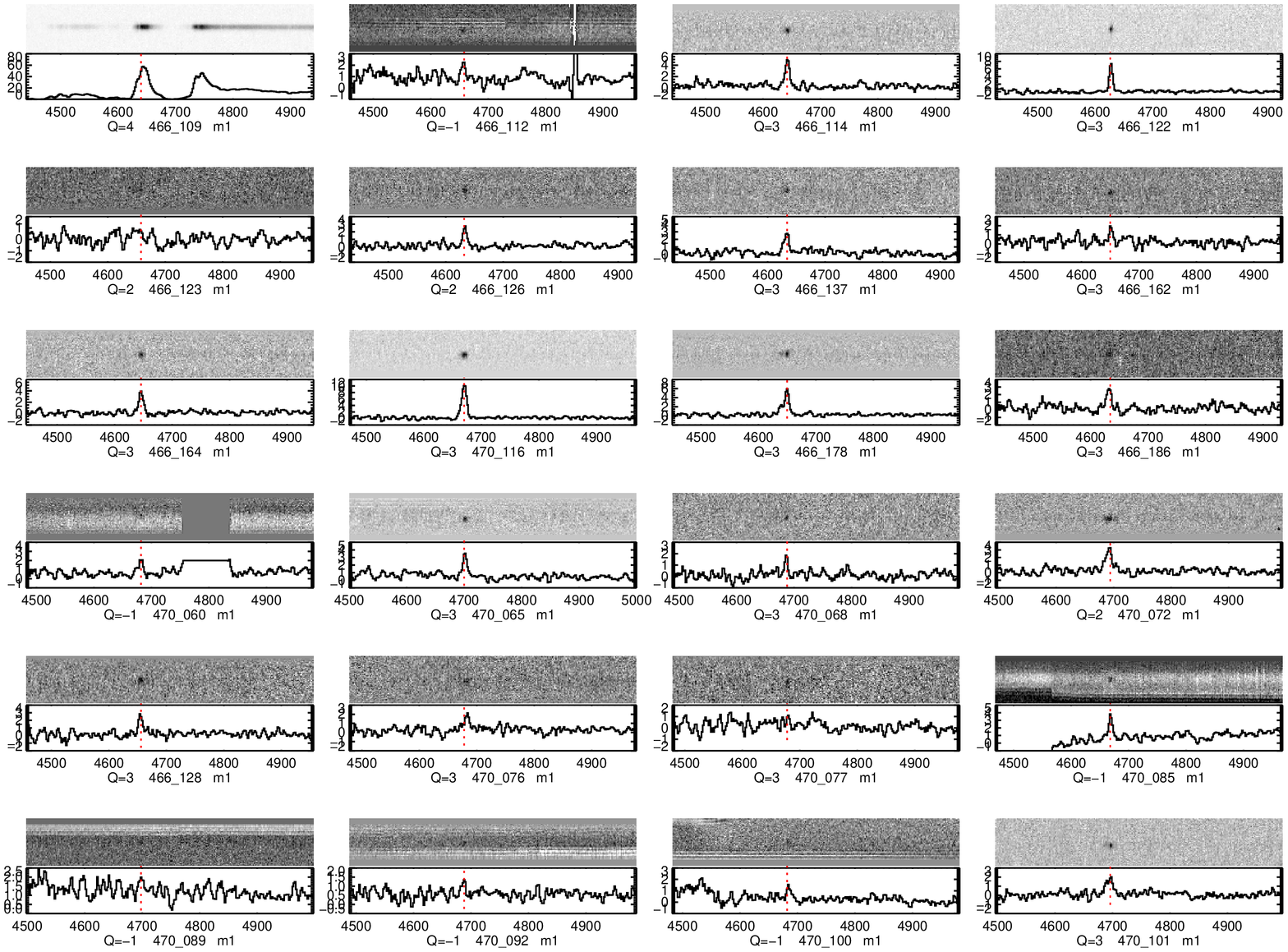} 
\caption{1-D and 2-D spectra of the spectroscopically confirmed LAEs (part 1). The spectra with best quality are marked as `Q=3', and negative Q value means the spectra are contaminated by nearby slits. The object name suffixes `m1' or `m2' imply the galaxy is confirmed with mask1 or mask2, respectively.   The unit of Y-axis is $10^{-18}$ erg cm$^{-2}$ s$^{-1}$ \AA$^{-1}$. For display only, the 1D spectra are smoothed with 3-pixel box.  }
\label{spec1}
\end{figure*}

 \begin{figure*}
\includegraphics[angle=270,width=\linewidth]{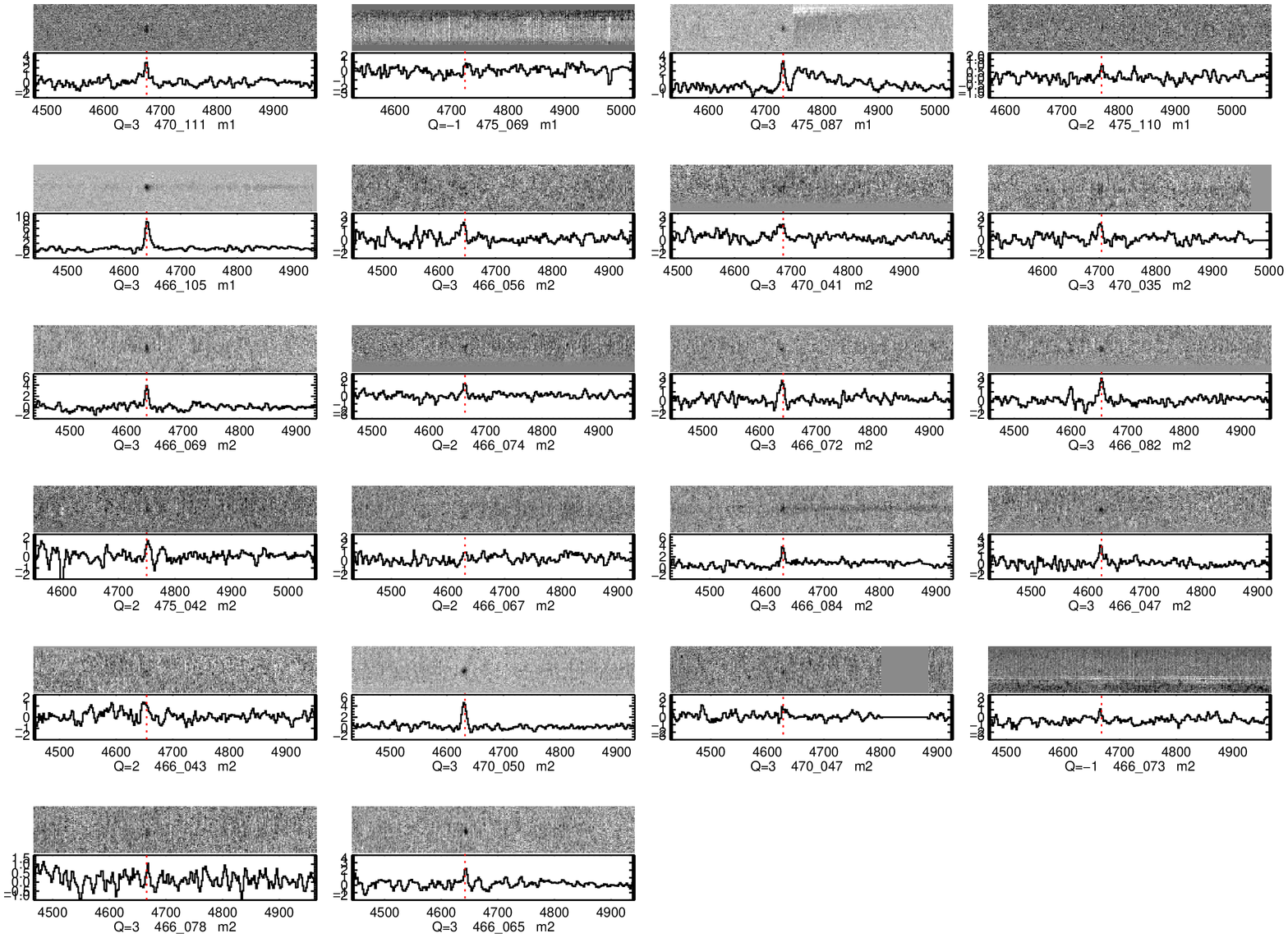} 
\caption{1D and 2D spectra of the spectroscopically confirmed LAEs (part 2). }
\label{spec2}
\end{figure*}

\subsection{Spectroscopic Confirmation of Emission-Line Galaxies}
\label{sec:specz}

Our spectroscopic data were obtained using the IMACS \citep{Dressler06} short 
camera ($f/2$, with a 27\arcmin .2 diameter field of view)  at the 6.5 m Magellan I Baade 
Telescope on 2013 November 27-28 (through Steward Observatory time, PI Zheng)
with the 300 line/mm grism. The 300 line/mm  grism has a $\lambda_{blaze}$ = 5000 \AA, 
and a resolution of 1.341 \AA\ pixel$^{-1}$ with a range of 3500-11000 \AA.  
Two multislit masks were  observed for 2.5 hrs and 3 hrs in 0.5 hr increments. The masks included $\sim$30\%
of the total emission line galaxies (81 in mask1 and 47 in mask2).
The masks have slit widths of 0.8 arcsec. The seeing during the observing period was $\sim$0.8--1.3\arcsec.  
 
The data were reduced using the IMACS version of the Carnegie Observatories System for MultiObject Spectroscopy (COSMOS) data 
reduction package\footnote{http://obs.carnegiescience.edu/Code/cosmos/Cookbook.html} with same steps introduced in Sec 2.2 of \citet{Zheng13}. 
In this step, COSMOS failed to extract 4 targets' spectra in each mask. Then we used our own idl program to combine the 2-d spectra in each frame with 
weighting factor of $\eta /$FWHM$^2/\sigma_{sky}^2$ as a function of transmission $\eta$ (relative scale to the standard star spectrum), seeing 
FWHM (from the spatial dispersion of the standard star spectrum), and background noise $\sigma_{sky}$ around the narrowband wavelength range. 

There are 46 spectroscopic targets confirmed as LAEs at $z=$  2.8--2.9 based on a single emission line at the expected  
wavelength range and spatial position. We determine their redshifts from the peak wavelength of their Ly$\alpha$ emission lines.  These LAEs 
are presented in Table \ref{tbl-1}, and their 1-d and 2-d spectra are plotted in Figure \ref{spec1} and Figure \ref{spec2}. 

The main low-z interlopers are expected to be [O\,{\sc ii}] emitters at z $\sim$ 0.25--0.27. Our spectroscopic survey cannot resolve 
the [O\,{\sc ii}]$\lambda\lambda$3727,3729 doublet, but its wavelength range covers the [O\,{\sc iii}], H$\beta$, and H$\alpha$ lines 
of the [O\,{\sc ii}] emitters at z $\sim$ 0.25--0.27. However, through the spectroscopic check, we find more kinds of low-z 
interlopers (even high-z interlopers) and fake emission line candidates in our sample.  We only find 2 [O\,{\sc ii}] emitters 
with significant detections of their [O\,{\sc iii}], H$\beta$, and H$\alpha$ lines.  2 Mg\,{\sc ii} emitters are found from the detections of their 
[O\,{\sc ii}]$\lambda$3727 and [O\,{\sc iii}]$\lambda$5007 lines.  3 [O\,{\sc vi}] emitters at z $\sim$ 3.4 are found via their significant break at the 
blue-end compared with the red-end of their Ly$\alpha$ lines. There are 8 targets with single emission line in the wavelength range of 5000--8000 
\AA\ and continuum at both end, therefore it is hard to determine their redshifts. It is also hard to determine the redshifts of several spectra with only continuum. 

In summary, we identify 46 (29 in mask1, and 17 in mask2; 25 in NB466, 17 in NB470, and 4 in NB475) LAEs at z$\sim$2.8--2.9, 
3 [O\,{\sc vi}] emitters at z $\sim$ 3.4, 8 possible LBGs (no lines, only continuum) and 12 low-z emitters (2 [O\,{\sc ii}] emitters at z$\sim$ 0.26, 
2 Mg\,{\sc ii} emitters at z$\sim$ 0.62, and 8 unknown galaxies with single-line plus both side continuum) from our 124 targets. Among the 45 
un-successful candidates, 3 are located in the CCD gaps, 17 are located in the bad spectral regions, and no line is found for the remaining candidates.

\begin{figure}
\includegraphics[angle=270,width=\linewidth]{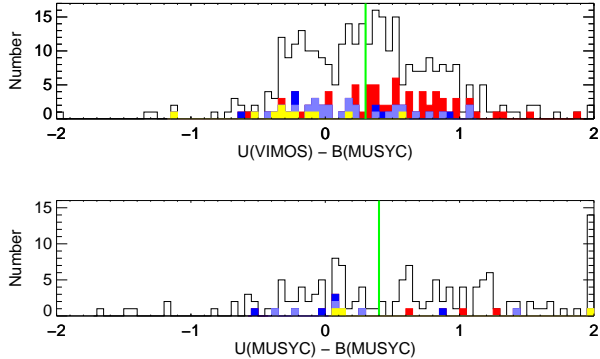} 
\caption{The $U$-$B$ color distributions of our sample. 
The black and red filled histograms show the $U$-$B$ color distributions of all emission line galaxies and 
the spectroscopically confirmed $z\sim$\,2.8 LAEs, respectively.
The filled histograms with other colors present targets not confirmed at $z\sim$\,2.8, 
including low-$z$ interlopers (yellow filled histogram), and unsuccessful candidates that are probably too
faint (light-blue filled histogram) or with marginal detections (blue filled histogram).
We choose $U_{\textsc{VIMOS}}$-$B_{\textsc{MUSYC}} \geq 0.3$ (corresponding to $U_{\textsc{MUSYC}}$-$B_{\textsc{MUSYC}} \geq 0.4$, 
green solid lines) as the selection criterion for LAEs at $z\sim2.8$.  }
\label{color}
\end{figure}

\begin{figure}
\includegraphics[angle=270,width=\linewidth]{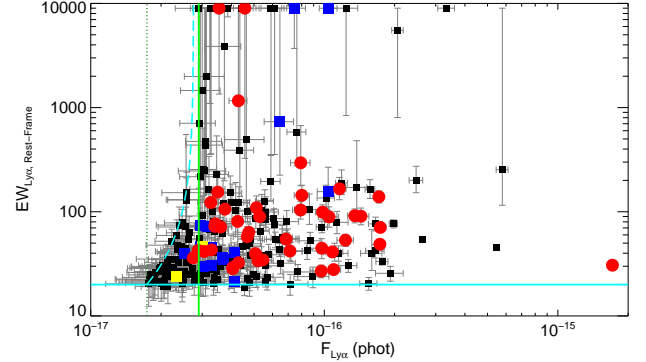} 
\caption{The distribution of \lya\ rest-frame EWs and \lya\ emission line fluxes for all LAE candidates from all three narrowbands 
(small black filled-squares), spectroscopically confirmed LAEs (red circles) and unconfirmed LAE candidates (blue 
filled-squares, including two low-z emitters marked with yellow filled-squares). The EW errors are calculated from Monte-Carlo 
simulations of both broad and narrowband data. The lines show the completeness limits of the survey. The dark-green dotted line 
shows the minimal  \lya\ emission line flux the survey can reach ($NB$ = 25 and $B$ = 26), and the cyan long-dashed line shows 
the LAEs with $NB$ = 25 and $B$ $\geq$ 26. The green vertical line and cyan horizontal curve show the complete LAE sample with 
$EW_r$ $\geq$ 20\AA\ and F$(Ly\alpha)$ $\geq$ 2.9 $\times$ 10$^{-17}$ erg cm$^{-2}$ s$^{-1}$.   }
\label{flyaew}
\end{figure}

\subsection{Ly$\alpha$ Galaxy Selection at z $\sim$ 2.8}
\label{sec:lyasel}

With the spectroscopic observation of selected emission line galaxies, we can explore the completeness and reliability for LAE selection 
at z$\sim$ 2.8. Because VIMOS-$U$ is much deeper than MUSYC-$U$, 
we use the VIMOS-$U$ band data for $\sim$66\% ELGs which are covered by VIMOS-$U$.
We find that the $U$-$B$ cut of  ($U$-$B$\,$\geq$\,0.3) $||$ ($U$-$B$\,$<$\,0.3 $\&$ $B/\sigma_B$\,$<$\,2) (for VIMOS-$U$ and MUSYC-$B$) is a relatively complete selection criteria 
(see Fig. \ref{color}). Through stellar synthesis model \citep[e.g., ][]{BC03},  ($U$-$B$ $\geq$ 0.3) would find galaxies at $z$\,$\sim$\,2.8 
with age above $\sim$1 Myr and metallicity above 0.4 $Z_\odot$, and larger age or metallicity or extinction would introduce larger $U$-$B$ value. 
Note  the profile of VIMOS-U is relatively redder than that of MUSYC-U, which would cause a difference of 0.1 for the color cut. 
There are 17 candidates with B-band non-detection and color $U$-$B$\,$<$\,0.3. Three of these galaxies are targeted, two are confirmed as $z$\,$\sim$\,2.8 LAEs
and the third one has no detection of any lines or continuum. These 17 candidates are included in the LAE sample at $z$\,$\sim$\,2.8.  
The B-band non-detection ($U$-$B$\,$<$\,0.3 $\&$ $B/\sigma_B$\,$<$\,2) is chosen as the supplementary of the color cut. The reliability of 
$U$-$B$ cut is 43/58 $\sim$ 74\% (58 LAE candidates are targeted, and 43 of them are confirmed). 
There are 3 low-$z$ interlopers and 1 possible $z$=2.34 galaxy among the 15 unconfirmed LAE candidates 
(see Table \ref{ecdet}), therefore the contamination fraction of our LAE sample is in the range of 7--26\%.
The completeness fraction is 43/50 $\sim$ 86\% (among the 50 emission line galaxies confirmed at $z\sim$ 2.8, 
which includes 4 galaxies confirmed from other surveys,  43 of them are previously selected as LAE candidates). 
So here we conclude our criterion -- the $U$-$B$ color cut as ($U$-$B$ $\geq$ 0.3) $||$ ($U$-$B$ $<$ 0.3 $\&$ $B/\sigma_B$ $<$ 2) -- 
for the LAE selection at $z\sim$ 2.8--2.9.  The line-excess plus color-cut selection method selects 257 LAE candidates in this 
survey (125 in NB466, 71 in NB470, and 61 in NB475).

We plot the distribution of \lya\ rest-frame EWs and \lya\ emission line fluxes for all LAE candidates selected above, and mark the spectroscopically confirmed 
and un-confirmed targets in Figure \ref{flyaew}. Obviously, our narrowband selection criteria would miss LAEs with large EWs in the line flux range of
 1.7$\times10^{-17}$ $\leq$ F$(Ly\alpha)$ $\leq$ 2.9 $\times10^{-17}$ erg cm$^{-2}$ s$^{-1}$ (above the cyan dashed line in Fig. \ref{flyaew}, which 
 is corresponding to LAEs with $NB$ $>$ 25 and $B$ - $NB$ $\geq$ 1). So in the following analysis, we only consider LAEs with F$(Ly\alpha)$ $\geq$ F$_{cut}$ = 2.9 
$\times10^{-17}$ erg cm$^{-2}$ s$^{-1}$ as the complete LAE sample at $z\sim$ 2.8--2.9 (unless specifically pointed out). This sample has 186 LAE candidates (97 in NB466, 59 in NB470, and 30 in NB475). In this sample, 55 LAE candidates are spectroscopically targeted and 41 of them are confirmed. 
In the following two sections, we match our complete LAE catalog with the \textit{GALEX} catalog and public spectroscopic catalogs to exclude the low-$z$ objects and objects confirmed at other redshifts. There are 2 LAE candidates detected by \textit{GALEX} as low-$z$ interlopers,  and 5 other LAE candidates with redshifts at $z<$ 2.8. 
Finally, the number of the complete LAE sample at $z\sim$ 2.8--2.9 is 179. 
In each field, the numbers are 96 from NB466, 55 from NB470, and 28 from NB475, respectively. 
Note that the NB466 probes overdense regions while the NB470 and NB475 observe blank fields.

\subsection{AGN and Low-z Interlopers Detected by {\it Chandra} and \it{GALEX}}
\label{sec:uvxdet}

By matching the \textit{Chandra} X-ray point source catalog of the 4\,Ms CDFS \citep{Xue11} and the four 250\,ks exposures of the ECDF-S 
\citep{Lehmer05}, we find 10 emission line candidates within $<$\,1\arcsec\ separations from the X-ray counterparts. Only 4 of these 
ELGs are LAE candidates, including our brightest emission line candidate which passed the LAE selection at $z\sim2.8$ and is 
 confirmed as a BAL-QSO at z=2.81 \citep{Szokoly04}. The remaining 6 ELGs do not pass the $U$-$B$ cut, and 
two of them are detected in all three narrow-bands and  confirmed as interlopers at $z=0.977$ (N470-147 as a Fe\,{\sc ii} emitter) and 
$z=0.68$ (N475-128 as a Mg\,{\sc ii} emitter). We have no spectroscopic information for the remaining galaxies. 

We next compare our list of emission-line galaxies to the catalog of UV sources detected by the {\it Galaxy Evolution Explorer} satellite
\citep[$GALEX$;][]{Christopher03}. $GALEX$ has conducted deep near-UV and far-UV ($m_{AB}$ $\sim$ 25) 
surveys of the ECDF-S region. The wavelength range covered by $GALEX$ are far beyond the Lyman 
break at $z\sim2.8$,  thus any LAE candidates listed in that catalog are most likely foreground contaminants. We find 18 emission line candidates 
within $<$\,1\arcsec\ separations, including two  also detected in X-ray. Only 2 candidates with $GALEX$ 
counterparts pass the LAE selection, of which one has a very large U-band error bar (N470-086, $U_m$ = 26.9$\pm$0.7), and 
the other one (N470-115) is confirmed as an [O\,{\sc ii}] emitter at $z=0.265$ \citep{Popesso09, Balestra10}. 
We list each of these X-ray bright and/or UV-bright sources in Table \ref{xdet}. Here we exclude the two NB470 LAE 
candidates with {\it GALEX} detections, but keep the three other LAEs detected by {\it Chandra} in the complete LAE sample.

\subsection{Matching with Public Spectroscopic Surveys}

The CDF-S is a field full of spectroscopic surveys from which we can take advantages of.  
ESO had collected the publicly available spectroscopic surveys in that 
field\footnote{http://www.eso.org/sci/activities/garching/projects/goods/Mas\\terSpectroscopy.html}, including VLT VIMOS 
and FORS2 spectroscopic surveys, i.e., \citet{Vanzella08} for ESO-GOODS/FORS2 survey,  
\citet{LeFevre04}, \citet{Popesso09} and \citet{Balestra10} for VLT-VIMOS Spectroscopic survey, and
\citet{LeFevre13, LeFevre15} for VLT-VIMOS Deep Survey (VVDS) and VIMOS Ultra-Deep Survey (VUDS). We compare the list of our candidates 
to the catalogs from these surveys. The matching procedure (matching radius of 1$\arcsec$) produces 30 coincidences. 10 of 
these have been confirmed at $z\sim$\,2.8--2.9, and have passed the LAE color-cut except one. There are 8 other ELGs passed our LAE color-cut, including one 
[O\,{\sc ii}] emitter at $z=0.265$ with {\it GALEX} detection, three low-z interlopers at $z=$ 0.416, 0.534 and 1.091, and four $z>2$ 
foreground galaxies (at $z=$ 2.737, 2.566, 2.343, and 2.339).  The matched emission line candidates are listed in 
Table \ref{ecdet}. Here we exclude these 7 interlopers, and leave 179 LAE candidates (96 in NB466, 55 in NB470, and 28 in NB475) in the complete LAE sample.

\section{Spectroscopic Results}
In this section we firstly introduce the spectroscopic calibrations and slit-loss estimations, then discuss the 
weak spectral features from the co-added spectrum of the confirmed LAEs. In the stacked spectrum, we 
only find a 4.5$\sigma$ detection of C\,{\sc iii}]$\lambda$1909 beside \lya, and find a velocity offset of $\sim$ 300 km~s$^{-1}$ 
between the two lines.


\begin{figure}
\includegraphics[angle=270,width=\linewidth]{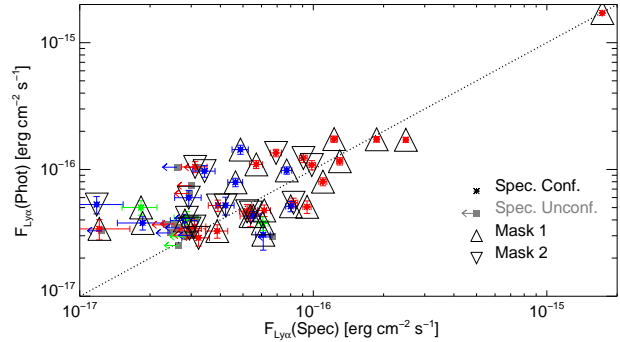} 
\caption{Comparison between the \lya\ fluxes calculated from the narrow-band photometric observation and those from the 
spectroscopic spectra. The grey squares and arrows mark the spectroscopically unconfirmed LAE candidates with the 
3-$\sigma$ upper limits of $F_{Ly\alpha}$(Spec). The triangles indicate spectra from mask1 or mask2. The 
sources from NB466, NB470, and NB475 are marked in red, blue and green colors, respectively. }
\label{fspevspho}
\end{figure}

\subsection{Spectroscopic Calibration}

The standard star UID1147 \citep{Pirzkal05} and the $z$\,$\sim$\,2.8 quasar \citep{Szokoly04} in each mask are used to 
calibrate the spectral flux. The 2-D and calibrated 1-D spectra are presented in Figure \ref{spec1} and Figure \ref{spec2}. 
The slit-losses are calculated by comparing the integrated narrow-band fluxes from the calibrated 1-D spectra and the 
photometric narrowband values. When integrating over the calibrated spectra, the narrowband filter profile is applied. 
We get an average slit-loss fraction of $\sim$46\%, with a  dispersion of 0.3 dex. The comparison of Ly$\alpha$ line fluxes 
integrated from spectroscopy and estimated from photometry (i.e., see Eq. A-1) is plotted in Figure \ref{fspevspho}.
Note the flux ratio dispersion here, which might be caused by several factors, such as slit losses (sensitive to the seeing, and 
the accurate position of the target in the slit) in spectroscopic observation, the nonuniform narrowband filter curve,  different 
spectral slopes (we assume a flat spectral slope in Eq. A-1), and the different IGM absorption at the blue-end of Ly$\alpha$ line 
at $z=$ 2--3.  Recently \citet{Momose16} reported that faint LAEs tend to have prominent diffuse \lya\ halos. 
Our multi-slit spectroscopic observation may miss the more extended \lya\ emission, therefore it is not strange to see the trend of increasing 
F$_{\rm Phot}$(Ly$\alpha$)/F$_{\rm Spec}$(Ly$\alpha$) ratio with the decreasing F$_{\rm Spec}$(Ly$\alpha$) line flux 
in Figure \ref{fspevspho}. We will explore these aspects in the next work with SED fitting on the $z=$ 2.8--2.9 LAEs, 
where we'll consider both narrowband and broad band images, and the nebular emission lines constrained from the 
spectroscopic observations (Zheng et al. in Prep.). The 3-$\sigma$ upper limits of the spectroscopic \lya\ fluxes of the 
un-confirmed LAEs are also presented in Figure \ref{fspevspho}. We notice that LAE N466-086 has no line detection in our 
spectroscopic observations, but was confirmed with a much deeper VUDS survey \citep{LeFevre15}. These imply that our 
spectroscopic survey is not deep enough to confirm all LAE targets.

\subsection{Stacked Spectrum}
\begin{figure}
\includegraphics[angle=270,width=\linewidth]{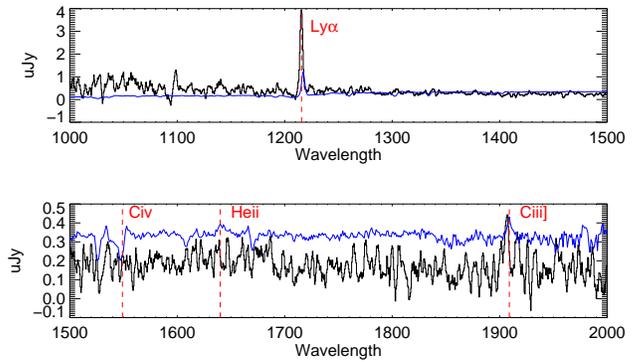} 
\caption{The median 1-D spectrum for the confirmed LAEs at $z\sim$\,2.8, in the rest-frame wavelength range of 1000--1500\AA\ (upper panel) and 
1500--2000\AA\ (lower panel). For comparison, the average spectrum of LBGs at $z\sim3$ \citep{Shapley03} is presented in blue. The \lya\ line and nebular lines such as 
C\,{\sc iv}, He\,{\sc ii}, and C\,{\sc iii}] are marked in red.
 }
\label{coaddavg0}
\end{figure}

 Co-adding the spectra will help us to explore weaker spectral features which are hidden in the noise of individual spectra. We apply
the median averaged stacking method introduced by \citet[][]{Shapley03}
to co-add the confirmed LAEs at $z\sim$\,2.8. There are 31 confirmed LAEs with good spectral quality (Q=3 in Fig. \ref{spec1} and \ref{spec2}). 
We use their \lya\ redshifts as the systematic redshifts and co-add their spectra without normalization. 
After masking out the sky-emission 
line regions, we obtain the median spectrum along the wavelength bins. The median spectrum is plotted in 
Figure \ref{coaddavg0}. The visible emission line features in the composite spectrum are the \lya\ and C\,{\sc iii}] lines only. 
There is no visible absorption line feature in the composite spectrum because of the high background fluctuation.
In the stacked spectrum, the \lya\ line has flux of 3.3$\pm$0.2 $\times$ 10$^{-17}$ erg cm$^{-2}$ s$^{-1}$, and rest-frame 
EW value of 41$\pm$2 \AA. The S/N of the only nebular line C\,{\sc iii}]$\lambda$1909 is 4.5$\sigma$. It has a line ratio 
to \lya\ of f(C\,{\sc iii}])/f(\lya) $\sim$ 1/8, and rest-frame EW value of 10.3$\pm$2.3 \AA. These EWs are consistent with 
the low mass and low luminosity gravitationally lensed C\,{\sc iii}] emitters at $z\sim$ 2 \citep{Stark14}.

\begin{figure}
\includegraphics[angle=270,width=\linewidth]{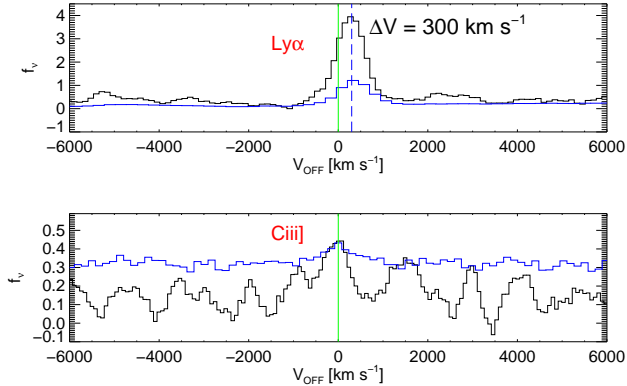} 
\caption{The outflow velocity estimated from the co-added spectrum. Bottom: the co-added spectrum at around C\,{\sc iii}]$\lambda$1909\AA, as a 
function of velocity offset relative to the systematic redshift from the peak of C\,{\sc iii}]; Top: same as bottom but around Ly$\alpha$$\lambda$1215.67\AA. 
The green vertical lines present $\Delta V$ = 0, and the blue dashed line marks the \lya\ peak in the systematic redshift frame, which is $\Delta V$ = 
300 km s$^{-1}$. 
For comparison, the average spectrum of LBGs at $z\sim3$ \citep{Shapley03} is presented in blue.  }
\label{coaddavg}
\end{figure}

From zoom in stamp of the stacked nebular line, the peak of C\,{\sc iii}] line is located at $\lambda$=1907\AA, with a \text{marginal} secondary peak at $\lambda\sim$ 1905\AA. 
Because we stack the spectra by setting their \lya\ peaks as their systematic redshifts, the relocation of C\,{\sc iii} doublet to rest-frame 1907,1909\AA\ implies
$\sim$300 km s$^{-1}$ red-shiftting of \lya\ line (see Fig. \ref{coaddavg}). This velocity offset, $\Delta V$ $\sim$ 300 km s$^{-1}$, is coincident with the \lya\ velocity offset from the average spectrum of LBGs at $z\sim$ 3 \citep[][see Fig. \ref{coaddavg} for the comparison]{Shapley03}, and agrees with the individual velocity offsets of the low mass and low luminosity gravitationally lensed C\,{\sc iii} emitters at $z\sim$ 2 \citep[60--450 km s$^{-1}$ with a mean of 320 km s$^{-1}$ between \lya\ and unresolved C\,{\sc iii} doublet, see ][]{Stark14}.
This velocity offset is also consistent with the individual measurements between \lya\ and [O\,{\sc ii}]/[O\,{\sc iii}] for LAEs at $z\sim 2$--3 \citep[e.g.,][]{McLinden11, Hashimoto13, Song14}.  This offset can be caused by outflows, while it can also be caused by the column density, the neutral gas, and the dust \citep[e.g.,][]{Verhamme06,Verhamme15}.

\begin{figure}
\includegraphics[angle=270,width=\linewidth]{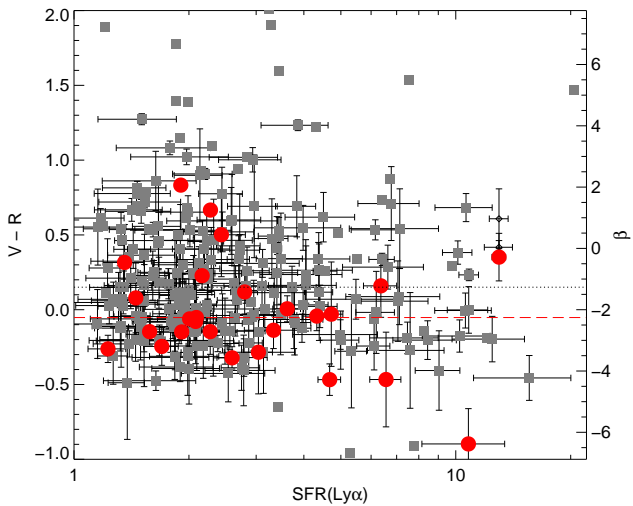} 
\caption{The $V$-$R$ color distribution as a function of SFR(Ly$\alpha$) for the LAE sample at $z$\,$=$\,2.8--2.9. The subsample used for the co-added spectrum
are marked with red filled circles. The average $V$-$R$ color of the whole LAE sample and the subsample are presented as the horizontal dotted and 
dashed lines, respectively.  The right axis shows the UV slope $\beta$ ($f_\lambda \propto \lambda^\beta$), which can be estimated from  $\beta$\,=\,4.88\,$\times$\,($V$-$R$)\,-\,2. 
Note that error-bars are presented only for LAEs with $\geq$3-$\sigma$ detections in both $V$ and $R$ band.  }
\label{vmr}
\end{figure}

Our stacked spectrum is steeper ($\beta$\,$\sim$\,-2.5, here $f_\lambda \propto \lambda^\beta$ ) than the average LBG spectrum ($\beta$\,$\sim$\,-2).  However, 
after checking the $V$-$R$ color distributions, the subsample used for spectra co-adding are bluer than the whole LAE sample on average (see Fig. \ref{vmr}). 
The subsample has an median and average $V$-$R$ value of -0.06 and -0.05, but with large standard deviation of 0.3. The whole LAE sample has an average 
$V$-$R$ value of $\langle V$-$R\rangle$ = 0.11$\pm$0.61. The UV slope $\beta$ can be estimated as $\beta$ = 4.88\,$\times$\,($V$-$R$)\,-\,2. Then the subsample has an average 
UV slope of $\beta$\,=\,-2.3, and the whole sample has $\beta$\,=\,-1.5, but both with very large errors.

The fluctuations in the continuum blue-ward of the \lya\ line are likely caused by the large scale structures reported at lower redshifts of $z\sim$ 2.56, 2.44, 
and 2.30 \citep[][see Fig. 1]{Popesso09,Balestra10}.  These structures at lower redshifts decrease the optical depth of the \lya\ forest in the 
spectra of $z$\,$\sim$\,2.8 LAEs, thus can be found from the stacked spectrum. Unfortunately, our individual spectra are not 
deep enough to resolve the \lya\ forest regions, which prevent us to map the three-dimensional hydrogen density at lower redshift (2.2 $< z <$ 2.8).


\section{Star-Formation Rates at $z$ $\sim$ 2.8}

The star-formation rate (SFR) is the key property of high-redshift star-forming galaxies. However, due to a variety of observational 
methods, different indicators of the SFRs exist \citep[see][]{Kennicutt98}. In this section, we explore the SFRs of 
LAEs from UV, \lya, and X-ray luminosities. These SFRs are compared and used to constrain the escape fractions of \lya\ and UV 
photons in LAEs at $z\sim$ 2.8.

\begin{figure}
\includegraphics[angle=270,width=\linewidth]{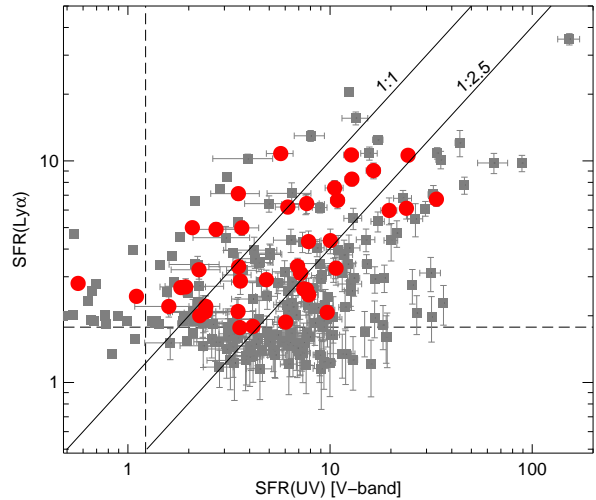} 
\caption{The comparison between SFRs calculated from V-band (UV flux density) and B band minus narrowband (\lya\ line flux) for the LAE 
sample at $z=$ 2.8--2.9. The spectroscopically confirmed LAEs are marked with red filled circles. The SFR(UV) from V-band limiting magnitude 
is presented as vertical dashed line. The horizontal dashed line presents F(\lya) = 2.9$\times$10$^{-17}$ erg cm$^{-2}$ s$^{-1}$.  Note that error-bars 
are presented only for LAEs with $\geq$ 3\,$\sigma$ detections in $V$ band. The average ratio of SFR(\lya)/SFR(UV) is $\sim$1/2.5 for the 
\lya\ and UV complete sample.  }
\label{sfruvsfrlya}
\end{figure}

Assume case-B recombination and the conversion from \citet{Kennicutt98}, 
the \lya\ flux from each narrow-band can be convert into the SFR$_{Ly\alpha}$, 
\begin{equation}
\textsc{SFR}(Ly\alpha) = 9.1 \times 10^{-43} L(Ly\alpha)\quad\textsc{   M}_\odot \textrm{ yr}^{-1}.
\label{sfrlyacorr}
\end{equation}
Similar conversion exists for SFR$_{UV}$ at $z\sim$2.8  from V-band flux density, 
 \begin{equation}
\textsc{SFR}(UV) = 1.15 \times 10^{-28} L_\nu \quad\quad\textsc{M}_\odot \textrm{ yr}^{-1},
\end{equation}
where $L_\nu$ (in erg s$^{-1}$ Hz$^{-1}$) represents the rest-frame mid-UV (1500-2800 \AA) continuum luminosity \citep{MD14}. 
Without the considering of resonant scattering of \lya\ photons and dust affection, galaxies which are optically thick to Lyman radiation should 
have \lya\ luminosities well correlated with their continuum flux. We plot the two quantities in Fig. \ref{sfruvsfrlya}. 
There is no significant correlation between SFR$_{Ly\alpha}$ and SFR$_{UV}$, but LAEs with brighter UV continuum tend to 
have less power to escape \lya\ photons (smaller SFR$_{Ly\alpha}$/SFR$_{UV}$ ratio in Fig. \ref{sfruvsfrlya}). 
This can be explained as UV brighter galaxies tend to be more massive and dusty, in which \lya\ photons are more difficult to escape.


The average SFR from  \lya\ and UV for all $z\sim$\,2.8 LAEs are 4.2 and 10.1 M$_\odot$/yr (excluding the matched 
X-ray sources in the complete LAE sample). In individual narrow-band images, the average  SFR(Ly$\alpha$) and 
SFR(UV) are [4.4, 11.2], [4.4, 8.9],  and [3.4, 9.0] M$_\odot$/yr, respectively. The average UV SFR from NB466 
is $\sim$25\% higher than that from the other two bands, and the average \lya\ SFR from NB475 is $\sim$23\% lower than 
that from the other two bands. Because the massive galaxies tend to have brighter UV radiation, this discrepancy would be the hints of
more massive galaxies and hence more clustered structures in NB466. The ratio of average SFR(\lya) to average SFR(UV) is 
about 0.4--0.5, which tells that dust and/or the radiative transfer of \lya\ photons exist in these galaxies.

Following \citet{Zheng12}, we stack the X-ray signal for the $z\sim$ 2.8 LAEs without X-ray detections. The LAE candidates located within 
6$\arcsec$ radius of nearby X-ray detections are excluded in the stacking.  We did not find any X-ray signal at $>$1.5 $\sigma$ significance in the soft, 
hard or total X-ray band. By stacking 40 LAEs located within 7-arcmin radius of the Chandra aim-point of 4Ms CDFS exposure, we get a 1-$\sigma$ 
upper limit of rest-frame L(2-10keV) $<$ 8 $\times$ 10$^{40}$ erg s$^{-1}$ with a total exposure time of 130 Ms. With the $L_X$--SFR$_X$ relations from 
\citet{Ranalli03}, \citet{Colbert04}, and \citet{Lehmer10},  we derive the 1-$\sigma$ upper limits of SFR$_X$ as 16, 50,  and 100 M$_\odot$/yr, 
respectively. 

The different $L_X$--SFR$_X$ relations are caused by different star-formation processes connected with X-ray emission. 
Most of the X-ray emissions connected with the star formations are emitted mainly by low-mass X-ray binaries (LMXBs) and high-mass X-ray 
binaries (HMXBs). With different evolution timescales (the age of the companion star), LMXBs are usually used to track the integrated star-formation of 
galaxies (i.e., the total stellar mass), while HMXBs are connected with the companion massive stars (sensitive to the instant SFR contributed by massive stars). 
The last two relations are calibrated from nearby normal galaxies \citep{Colbert04} and luminous infrared galaxies 
\citep{Lehmer10}. These galaxies are older and more dusty than typical star-forming galaxies, thus more sensitive to the 
low-mass X-ray binaries (LMXBs). The relation in \citet{Ranalli03} is calibrated in 
 a sample of local and high-z star-forming galaxies with the total SFR -- FIR relation from \citet{Kennicutt98} and total 
 SFR -- radio relation from Condon (1992), assuming X-ray SFRs from HMXBs. High-z LAEs are known as less massive star-forming 
 galaxies, thus we choose the X-ray-SFR relation in \citet{Ranalli03}. Therefore the SFR upper limit calculated from the average X-ray signal of our 
 sample is 16 M$_\odot$/yr. 
 
 Considering the penetration of X-ray photons in a galaxy (the typical column density in a galaxy is 
 only about 10$^{21}$ cm$^{-2}$, therefore the X-ray absorption can be ignored), the SFR$_X$ upper limit can be set as the 
 upper limit of intrinsic (unobscured) SFR. Compared to the average SFRs from \lya\ and UV estimated above, 
we can estimate that the \lya\ escape fraction is in the range of 25\% $\lesssim$ $f^{Ly\alpha}_{\rm ESC}$ $\lesssim$ 40\% 
(SFR$_{Ly\alpha}$/SFR$_X$ $\lesssim$ $f^{Ly\alpha}_{\rm ESC}$ $\lesssim$ SFR$_{Ly\alpha}$/SFR$_{UV}$), and the 
escape fraction of UV continuum photons is $f^{\rm UV,cont}_{\rm ESC}$  $\gtrsim$ 62\% ($f^{\rm UV,cont}_{\rm ESC}$ 
$>$ SFR$_{UV}$/SFR$_X$). The \lya\ escape fraction estimated here is consistent with the far-infrared stacking analysis 
by \citet{Wardlow14} with LAEs at $z\sim$ 2--3, and by \citet{Kusakabe15} with a larger sample of LAEs at $z\sim$ 2.2.


\section{\lya\ luminosity function at $z\sim$ 2.8--2.9}

The \lya\ luminosity functions (LFs) and \lya\ EW distributions
are fundamental observational quantities of \lya\ emitter galaxies.  
 Through a Monte Carlo approach with the considerations of the prior known \lya\ LF and EW distributions, 
the narrowband filter profile, and the simulated observational uncertainties, we find that 
the observational uncertainties and selection processes keep the shape of flux distribution, but
significantly boost the EW distribution to the high-value end (See Appendix Sec. A-3 and Fig. \ref{ewdist2}).
Therefore in this section we only measure and compare the \lya\ LFs following the methods introduced in 
\citet{Zheng13,Zheng14}. 
We introduce the \lya\ LF of z $\sim$ 2.8--2.9 LAEs in the ECDFS in Section \ref{sec:lyalfz28}. 
We also compare our \lya\ LFs with those of other surveys at $z\sim$ 2.1 and $z\sim$ 3.1 \citep{Ciardullo12, Ouchi08}.
Finally, we explore the evolution of the \lya\ LFs and the \lya\ photon densities over a large redshift range in Section \ref{sec:lyaevo}.

 \begin{figure}
\includegraphics[angle=270,width=\linewidth]{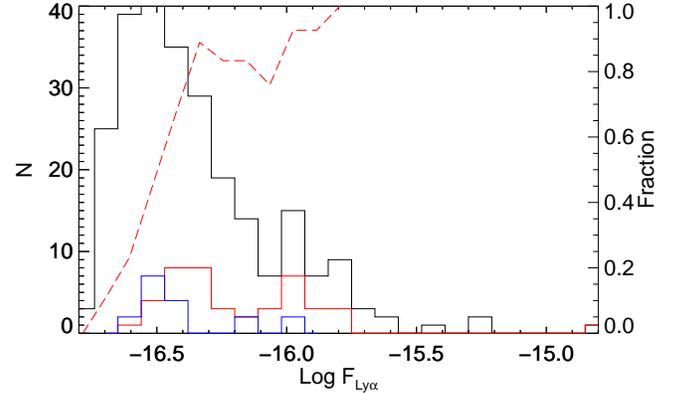} 
\caption{
Distributions of \lya\ fluxes for all LAE candidates (black empty histogram), spectroscopically confirmed LAEs (red histogram) and 
unconfirmed objects (blue histogram). The red dashed line shows the smoothed spectroscopic success fraction. }
\label{succ}
\end{figure}

\subsection{\lya\ LF at $z$ = 2.8--2.9 in ECDFS }
\label{sec:lyalfz28}

We use the $V/V_{max}$ method \citep[e.g.,][]{Dawson07, Zheng13} to calculate the \lya\ luminosity function. 
$V_{max}$ is the comoving volume where the source could be selected by our survey. We calculate $V_{max}$ 
for each confirmed LAE. Because of the limited redshift range of LAE survey, object with same luminosity only shows 
a decrease in 0.03 dex on its line flux in the NB466 and NB475 images, thus $V_{max}$ estimated here is nearly 
equal to our survey volume. The distortion caused by the profile of the narrowband filter is considered statistically. 
We mock the observing and selecting processes in Appendix, and find little effect on the recovery 
of \lya\ LF (see Fig. \ref{ewdist2}). The formula below is used to measure our \lya\ luminosity function:
\begin{equation}
\Phi(L)d L = \sum_{L-d L/2 \le Li \leq L+d L/2} \frac{1}{V_{max}(L_i)\times f_{comp}(L_i)},
\end{equation}
here V$_{max}(L_i)$ is the maximum volume in which LAEs with \lya\ luminosity L$_i$ can be found in our 
narrowband surveys, and $f_{comp}(L_i)$ is the completeness fraction for LAEs with \lya\ luminosity L$_i$.  
We use the completeness fraction of our narrowband detected objects (estimated from the exponential 
distribution in Fig. \ref{nbsel}) as an approximation, which is a function of narrowband magnitudes. The 
completeness fraction is $\sim$\,60\% at NB = 25 mag, corresponding to a \lya\ luminosity of 
log(L$_{Ly\alpha}$) = 42.3 with no continuum. The spectroscopic success fraction is also a function 
of \lya\ luminosity. It reaches to $\sim$100\% for bright LAEs, while it declines to 10\% or even below that 
for faint LAEs (see Fig. \ref{succ}). Ideally, the spectroscopic success fraction agrees with the 
success fraction of our LAE selection. However, we should notice that the LAE selection is not complete 
due to the deviation of the narrowband filter profile from the ideal 'top-hat' profile and the observational 
uncertainties (see Appendix). The correction on the spectroscopic success fraction is not needed because 
the observing and selecting process can not change the shape of \lya\ LF. Therefore people usually do not 
consider the spectroscopic success fraction in each luminosity bin \citep[e.g.,][]{Ouchi08, Ciardullo12}.

\begin{figure}
\includegraphics[angle=270,width=\linewidth]{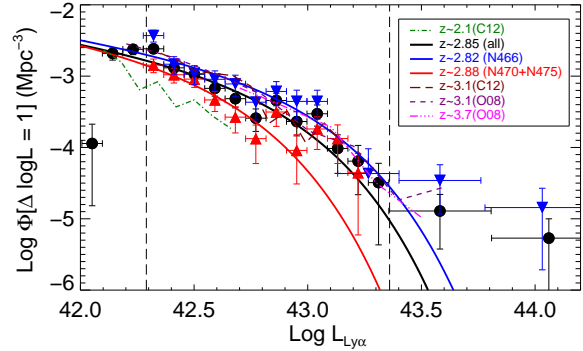} 
\caption{The \lya\ luminosity function of LAE galaxies at $z$\,=\,2.8--2.9. The sample is corrected for 
photometric selection completeness but not for spectroscopic success fraction. The \lya\ LFs between the 
two vertical dashed lines are selected to fit the Schechter function. The fitting results of $z\sim$ 2.8 LAEs 
are plotted here in solid curves, and the parameter contours are presented in Figure \ref{lyalflphi}.
The \lya\ LFs in the two subsamples (N466-only and N470+N475) are presented in blue and red triangles, respectively.
For comparison, we also plot the \lya\ LF of LAE survey at $z$ = 2.1 \citep[dark-green dot-dashed line for][]{Ciardullo12}, 
$z$ = 3.1 \citep[brown long-dashed line and purple dashed line for][respectively]{Ciardullo12, Ouchi08}, and 
$z$ = 3.7 \citep[magenta dash-dot-dot line for][]{Ouchi08}.  
}
\label{lyalf}
\end{figure}

\begin{figure}
\includegraphics[angle=270,width=\linewidth]{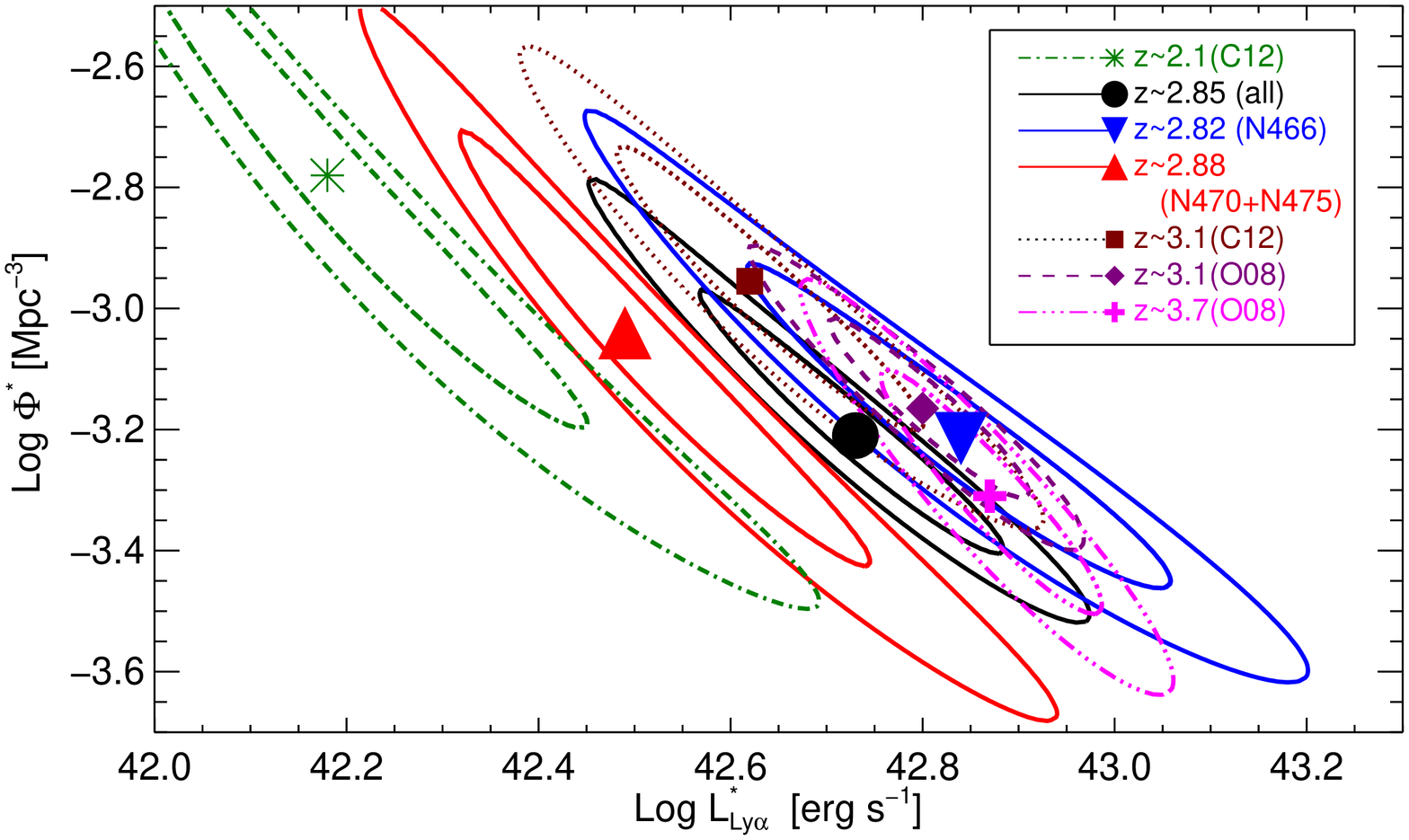} 
\caption{Contours of the parameters L$^*$ and $\Phi^*$ for our $z$ = 2.8--2.9 sample. The two contours 
denote the confidence level of 68 and 90 percent ($\Delta\chi^2$ = 2.3 and 4.6). The contours of the \lya\ LF 
parameters of the two subsamples (N466-only and N470+N475) are shown in blue and red, respectively.  
For comparison, we also plot the \lya\ LFs of LAE surveys at $z$ = 2.1 \citep{Ciardullo12}, $z$ = 3.1 
\citep{Ciardullo12, Ouchi08}, and $z$ = 3.7 \citep{Ouchi08} with the same colors as in Figure \ref{lyalf}. 
Note that all best-fitting parameters are obtained by fixing the faint-end slope $\alpha$ = -1.5. }
\label{lyalflphi}
\end{figure}

The derived \lya\ LF of LAEs at $z\sim$ 2.8 in the ECDFS is shown in Figure \ref{lyalf}. Following \citet{MR04}, 
we fit the \lya\ LF with a Schechter function:
\begin{equation}
\Phi(L)d L = \frac{\Phi^*}{L^*}\left (\frac{L}{L^*}\right )^\alpha \exp\left (-\frac{L}{L^*}\right ) d L .
\label{eqn:schechter}
\end{equation}
We use the IDL program \textit{mpfit} to fit the Schechter function with the $\chi^2$ statistics 
($\chi^2$ = $\sum_{i=1}^{N}(\Phi_i-\Phi_{mod})^2/Err_\Phi^2$). The fitting is applied only to the complete 
LAE sample without X-ray detection, which is marked as two dashed lines in Figure \ref{lyalf} (\lya\ luminosity 
range of 42.29\,$\leq$\,log$_{10}$(L$_{Ly\alpha}$)\,$\leq$\,43.36). 
We ignore the photometric errors of luminosities in the fitting, as we divide
our sample into bin-size of 0.09 dex, which corresponds to a $\sim$5-$\sigma$ detection in the faintest luminosity bin. 
Thus the photometric errors primarily affect our faintest bin. We find the best-fit parameters of 
log$_{10}$($L^*$) = 42.73$\pm$0.08 and log$_{10}$($\Phi^*$) = -3.21$\pm$0.11 ($\chi^2/dof$ = 18.5/10) with 
fixed $\alpha$\,=\,-1.5 \citep[same $\alpha$ as][]{Zheng13}.
When changing the fixed faint-end slope to $\alpha$\,=\,-1.65, we get nearly same best-fit parameters of 
log$_{10}$($L^*$) = 42.80$\pm$0.09 and log$_{10}$($\Phi^*$) = -3.33$\pm$0.12 ($\chi^2/dof$ = 16.7/10).  
Our results are within the 1-$\sigma$ range of $L^*$, but about 0.1--0.2 dex lower on $\Phi^*$ when compared 
to the \lya\ LFs at $z\sim$ 3.1 \citep{Ouchi08,Ciardullo12}. The Schechter function fitting results are presented in 
Tab. \ref{lyalfs}, and the contours of the fitting parameters are plotted in Figure \ref{lyalflphi}.

The faint-end slope in the LF has important implications for questions on galaxy formation and cosmic 
reionization \citep[e.g., ][]{Dressler15}. However, the complete LAE sample is not deep enough to robustly 
constrain the faint-end slope here. The \lya\ EWs are thought being independent of \lya\ luminosity 
\citep{Nilsson09, Zheng14}, thus we can include the fainter luminosity bins to fit the 
faint-end slope (it should be reminded that the LAEs in the  fainter luminosity bins are incomplete
as LAEs with larger EWs may be missed). By fitting in the \lya\ luminosity range of 42.1 $\leq$ 
Lg(L$_{Ly\alpha}$) $\leq$ 43.36, we find the best-fit parameters of the Schechter function as 
log$_{10}$($L^*$) = 43.1$\pm0.3$, log$_{10}$($\Phi^*$) = -4.0$\pm0.6$, and $\alpha$ = -2.1$\pm$0.3  ($\chi^2/dof$ = 15.7/11).
Although the large error on $\alpha$, the value of $\alpha$ = -2.1$\pm$0.3 is consistent with the much deeper 
blind search of LAEs at $z\sim$ 5.7 by \citet{Dressler15}. However, we should note that the increasing 
unsuccessful spectroscopic fraction and in-completeness photometric fraction toward the faint \lya\ luminosity 
bins may flatten the faint-end slope and introduce systematic uncertainties to $\alpha$.

\subsection{The Evolutions of \lya\ Luminosity Function and \lya\ Density}
\label{sec:lyaevo}

 We check the evolutions of \lya\ LF and \lya\ photon density over a large redshift range in this section.
The \lya\ LFs and fitting results at $z\sim$ 2.1, 2.8, 3.1 and 3.7 are  plotted in Figure \ref{lyalf} and Figure \ref{lyalflphi}. 
Our results show that the \lya\ LF at $z\sim$ 2.8 locates between the  \lya\ LFs at $z\sim$ 3.1 and $z\sim$ 2.1, 
with a $>$\,3\,$\sigma$ significant difference compared with that at $z\sim$ 2.1, and a $\sim$ 1--2\,$\sigma$ 
difference compared with that at $z \geq$ 3.1. However, the \lya\ LF at $z\sim$ 2.8 in NB466 and that 
in NB470+NB475 show a difference of $\gtrsim$ 2$\sigma$. 
 This could be caused by the overdense regions in the NB466 field, as NB466 observes the overdense regions while 
NB470 and NB475 observe the blank fields. We describe the whole overdense regions in Sec. \ref{sec:lss}, and compare the \lya\ LFs 
in overdense and blank fields in Sec \ref{sec:lyalfdiff} in details. 
The \lya\ LFs at $z\sim$ 2.1 and 3.1 come from blank fields.  
Here we find that the evolution of \lya\ LF in the blank fields at $z\sim$ 2.1, 2.8 and 3.1 agrees well with the general galaxy evolution.

 \begin{figure}
\includegraphics[angle=270,width=\linewidth]{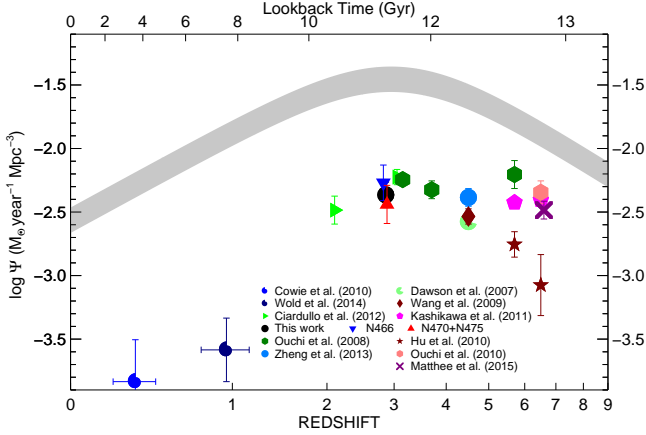} 
\caption{Measurements of the observed SFRDs of LAEs as a function of redshift. 
Each point is derived from the integrated \lya\ luminosity function down to 0.03$L^*$, 
and for consistency, each assumes a faint-end slope of $\alpha$ = -1.5. The black filled circle 
represents the SFRD of our LAE survey at $z=$ 2.8--2.9. We mark the SFRDs in the two subsamples
as a blue upside-down triangle (in NB466, the overdense field) and a red triangle (in NB470+NB475, the blank field).
The shaded area displays the observed SFRDs from UV summarized by \citet{MD14}. 
At $z$ $<$ 5, the \lya\ and UV-based measurements of the SFRDs are similar, while at $z$ $>$ 5, there are large scattering 
measurements of the \lya-based SFRDs.
}
\label{sfrds}
\end{figure}


 \begin{figure}
\includegraphics[angle=270,width=\linewidth]{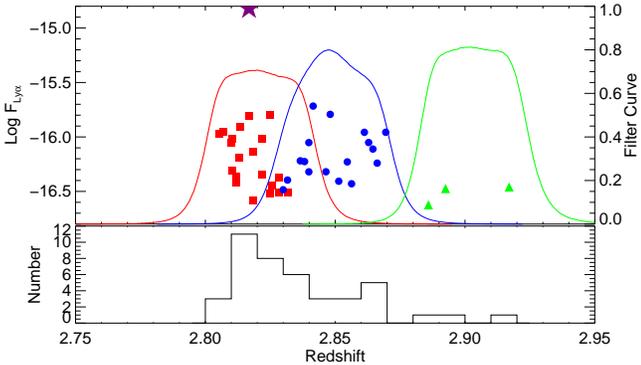} 
\caption{Top: The photometric \lya\ fluxes as a function of redshift for the spectroscopically confirmed LAEs, and the narrowband transmission curves with which the LAE
candidates are selected; Bottom: redshift distribution of all confirmed LAEs at $z$ = 2.8--2.9. The purple star is the BAL-QSO at z=2.81. }
\label{zdist}
\end{figure}

\begin{figure}
\includegraphics[angle=0,width=\linewidth]{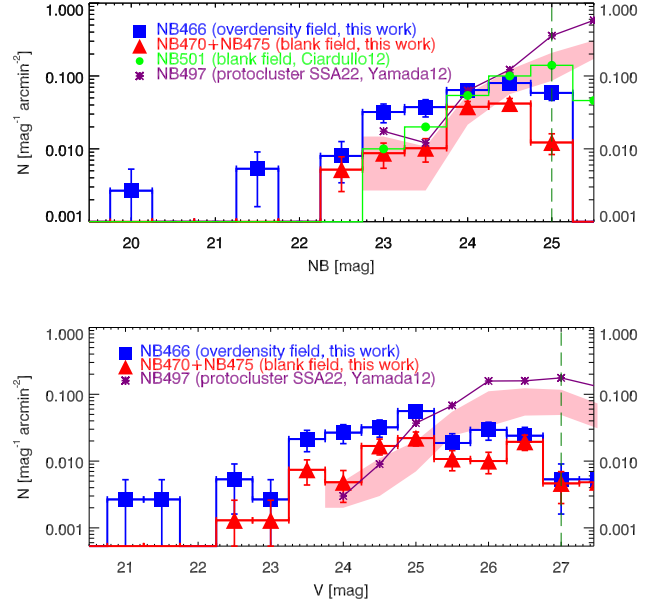}
\caption{
The Narrowband (top panel) and V-band (bottom panel) magnitude distributions of LAE surveys at $z\sim$ 3. 
LAEs from NB466 are marked with blue squares and histograms, and LAEs from NB470+NB475 are marked with red triangles and histograms.
The green histogram represents LAEs in same field but at $z=3.1$ from \citet{Ciardullo12}. The purple lines represent SSA22-Sb1 which has the highest density of LAEs at $z=3.1$ in SSA22 field.  The pink poly-filled region shows the distributions in SDF, SXDS, and GOODS-N fields at $z=3.1$ in \citet{Yamada12}. We scale all the surface densities here to match the survey volume of NB475. The vertical dashed lines in dark-green color show the narrowband and $V$ band depths of our survey. Because the depths of our narrowband images are shallower than those of other surveys, LAEs with faint NB or $V$ is less than those of other surveys.}     
\label{nbcts}
\end{figure}

 \begin{figure*}
\includegraphics[angle=0,width=\linewidth]{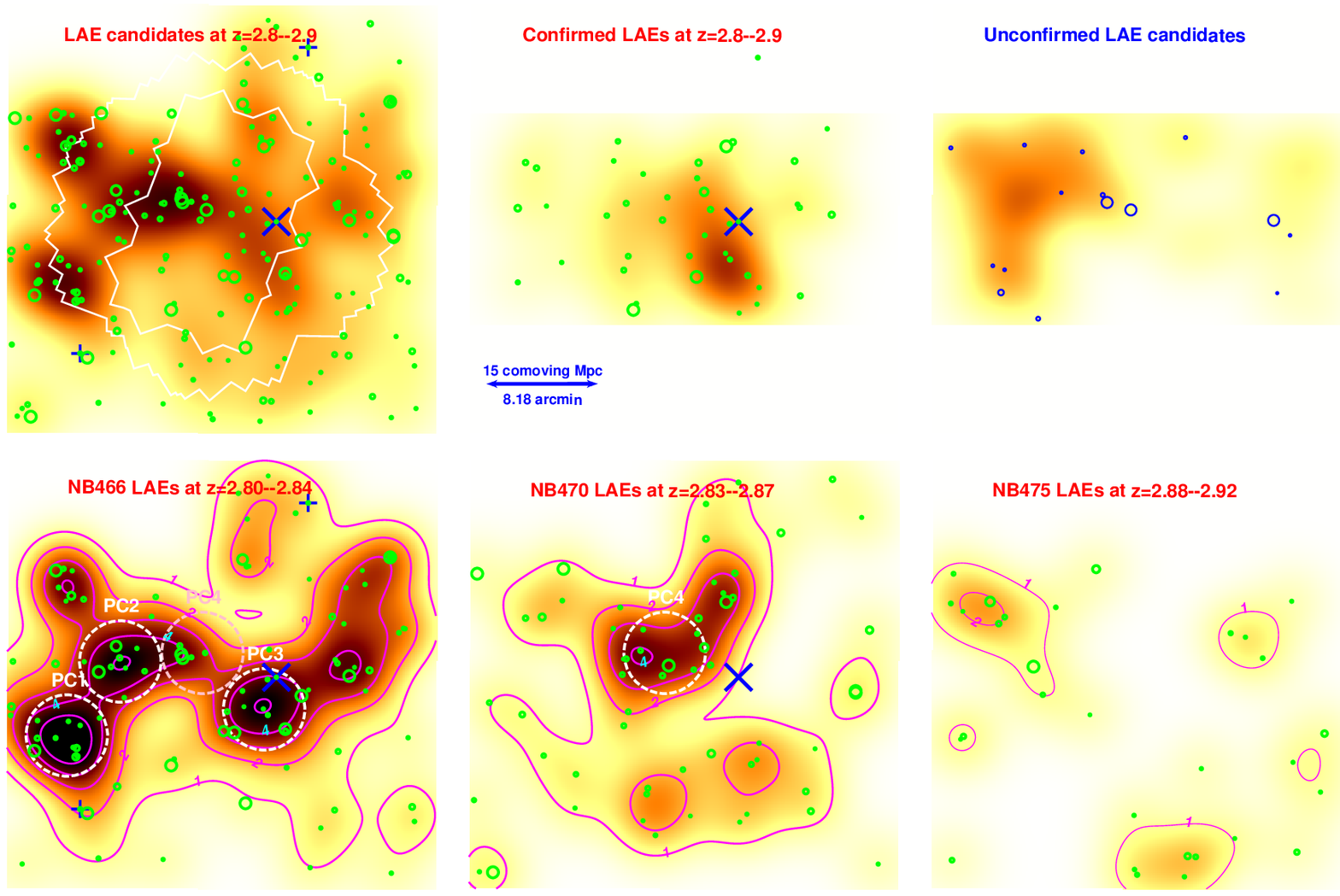} 
\caption{
The distributions of LAEs on the sky, for all LAE candidates (upper left), confirmed LAEs (upper middle),  
unconfirmed LAE candidates (upper right), and LAE candidates in each narrowband (lower panels).  The size of the 
circle is proportional to their \lya\ EWs. The images are smoothed with FWHM=5$'$ gaussian, and the contour levels 
in the bottom images are 1, 2, and 4 times of the average volume density of the complete sample in all three narrowband 
images (top-left panel). The BAL-QSO are marked as blue 'X', and the other X-ray detected LAEs are marked in blue crosses. 
There are significant over dense and void regions of $z\sim$\,2.8 LAEs in the three narrowband images. In particular, the 
overdensity in NB466 is dominated by a large-scale structure composed of three protoclusters (PC1, PC2, and PC3). Another 
protocluster (PC4) is found in NB470. The protoclusters PC1, PC2, PC3, and PC4 contain 12, 9, 11, 9 LAEs, respectively, 
within a radius of 3 arcmin (corresponding to a comoving volume of (15 Mpc)$^3$). }
\label{spatial}
\end{figure*}

With the \lya\ LFs of LAEs at different redshifts, we can explore the cosmic SFR density \citep[SFRD, ][]{Lilly96, Madau96} 
from LAEs as a function of redshift. With Eq. \ref{sfrlyacorr}, the SFRD of LAEs can be converted from the integrated \lya\ 
luminosity function over luminosity, which is log$_{10}$($\Psi_{Ly\alpha} [\textrm{M}_\odot \textrm{yr}^{-1} \textrm{Mpc}^{-3}]$) 
= log$_{10}$($\Gamma(\alpha+2,L_{min}/L^*)$\,$L^*\Phi^*$)\,-\,42.04 by integrating the Schechter function in 
Eq. \ref{eqn:schechter} over $L$. Although the value of L$_{min}$ is different for different LAE surveys, the incomplete 
gamma function $\Gamma(\alpha+2,L_{min}/L^*)$ would not change too much with a fixed $\alpha$ = -1.5 \citep[cf.,][]{Zheng13}. 
For consistency, here we apply $L_{min}$ = 0.03$L^*$, and $\Gamma(\alpha+2,0.03)$ = 0.8 $\times$ $\Gamma(\alpha+2, 0)$ = 1.43 
when $\alpha$ = -1.5. With the parameters in Figure \ref{lyalflphi}, the SFRD from \lya\ photons of LAE surveys are 
log$_{10}$($\Psi_{Ly\alpha}(z)$) = [-2.49$\pm$0.11, -2.37$\pm$0.10, -2.25$\pm$0.06, -2.33$\pm$0.12] at $z\sim$ [2.1, 2.8, 3.1, 3.7]. 
In the redshift range of 2--4, the \lya-based SFRD peaks at $z\sim$ 3.1, while it can also be explained as flat within their 1-$\sigma$ errors.

The SFRDs from \lya\ and UV are compared in Figure \ref{sfrds}. In this plot, the SFRDs from \lya\ are calculated from LAE 
surveys over redshift range of 0.3--6.6, including low-redshift ($z<$ 2) LAEs from {\it GALEX} slitless UV spectra 
\citep{Cowie10,Wold14}, and high-redshift ($z>$ 2) narrowband surveys \citep{Ciardullo12,Ouchi08,Ouchi10,Zheng13,
Dawson07,Wang09,Ka11,Hu10,Ouchi10,Matthee15}. The SFRD from rest-frame UV measurements are extracted from 
Table 1 of \citet{MD14}. The SFRDs from \lya\ are not corrected for the IGM absorption on \lya\ lines.

The difference between the \lya\ and UV-based SFRDs tells the evolution of dust and the radiative transfer of \lya\ photons 
over redshifts. At $z<$ 2, the ratio of \lya-SFRD and UV-SFRD is $\lesssim$ 0.03, while at $z>$ 2 the ratio increases to 0.1 
at $z\sim$ 4.5. It shows a large scatter at $z\gtrsim$ 5.7.  The observed SFRDs from \lya\ are nearly constant in the redshift 
range 2.8\,$\leq$\,$z$\,$\leq$\,5.7, while the UV SFRDs decrease over the same redshift range. This is consistent with the 
conclusion in \citet{Stark10} which shows that the prevalence of strong \lya\ emission increases moderately from  $z\sim$ 3 
to  $z\sim$ 6. The difference between the \lya\ and UV-based SFRDs can be linked to the global \lya\ escape 
fraction\footnote{The previous \lya\ escape fraction is for \lya\ galaxy itself, here the global \lya\ escape fraction takes into 
account the galaxies with little-to-no \lya\ emissions, which can't be selected from the narrowband LAE surveys.}\citep[c.f.,][]{Hayes11}, however, care should be taken when correcting the IGM absorption on \lya\ lines and correcting the completeness of different LAE surveys.

\section{A Large-Scale Structure at $z$ $\sim$ 2.8}
\label{sec:lss}

In this section, we confirm a large-scale structure at $z\simeq$2.8 from LAEs in NB466, which is indicated by a density peak reported in previous ESO/GOODS spectroscopic programs \citep{Balestra10, LeFevre15}. 
The number of LAEs in NB466 is larger than the total number of LAEs in NB470 and NB475, which implies that NB466 is an overdense field. 
We explore how significant this overdensity is in Section \ref{sec:spatial}, what this large-scale structure at $z\sim$ 2.8 will evolve into at $z\sim$ 0 in Section \ref{sec:protocluster}, and how the overdense environment would affect the physical properties of LAEs, e.g., the \lya\ luminosity functions (Section \ref{sec:lyalfdiff}), the distribution of \lya\ EWs and the distribution of colors (Section \ref{sec:protoclusterphy}).

\subsection{The Overdensity and Projected Distributions of LAEs at $z\sim$ 2.8}
\label{sec:spatial}

The number of LAEs in each field can be used to estimate the overdensity of LAEs at $z\sim$ 2.8. 
In the complete LAE sample, 96, 55 and 28 LAEs are selected with NB466, NB470 and NB475 filters, respectively.
There is an overlap along the line-of-sight between NB466 and NB470 (see the transmission curves in Fig. \ref{f1}). Hence, the effective volumes of the first two narrow bands are $17\%$ and $14\%$ smaller than that of NB475. After correcting for the different volumes, the average LAE surface densities become [0.13, 0.07, 0.03] arcmin$^{-2}$ and the volume densities become [8.97, 5.09, 2.23]$\times$ 10$^{-4}$ Mpc$^{-3}$ for the three narrowbands. The high density in NB466 is consistent with the peak in the redshift distribution of galaxies close to $z\sim$2.8 (Fig. \ref{f1}), which was reported in the previous work. Therefore, NB466 is an overdense field, and the other two regions NB470+NB475 can be treated as a general blank field. The density excess in NB466 is $\delta_{LAE}$ = ($\delta N/N_0$)$_{{LAE}}$ = 1.51$\pm$0.21 compared with the NB470+NB475 field, which is as significant as the protocluster SSA22 in the 647 arcmin$^2$ Sb1 field \citep[$\delta_{{LAE}}$ = 1.13,][]{Yamada12}. The mass fluctuation estimated from the standard $\Lambda$CDM model with the linear approximation is $\sigma_{mass}$ = 0.126 at $z=$ 2.8. 
Assuming a linear bias of b=2 for LAEs \citep{Gawiser07}, the density excess in NB466 is $\sim$  6.0$\pm$0.8 $\sigma_{LAE}$, which indicates that the NB466 filter is detecting a very rare high-density region.

Next, we compare our survey with the other two narrowband surveys at a close redshift $z\sim$3.1 in \citet{Yamada12} and \citet{Ciardullo12}.
\citet{Yamada12} presented the results of deep narrowband surveys of LAEs at $z\sim$3.1 in the 1.38 deg$^2$ SSA22 field, and in several
 blank fields with a total area of 1.04 deg$^2$. \citet{Ciardullo12} carried out a survey for $z$\,=\,3.1 LAEs also in the ECDFS field. Fig. \ref{nbcts} shows the magnitude
distributions both in the narrowbands and in V-band of our survey, in comparison with those at $z\sim$3.1.
It allows us to compare the number densities in different fields. Our narrowband images are $\sim$1\,mag shallower than in SSA22 (NB497), and $\sim$0.5\,mag shallower than $z\sim$3.1 ECDFS images (NB501), which leads to a smaller number of faint LAEs in our survey.
The magnitude distributions become flatter after $m_{NB}\sim 25$ in our survey, indicating that it is incomplete at $m_{NB}\geq 25$.
Therefore, we only consider the bright LAEs ($m_{NB}\leq 24.5$) in the density comparison.  Because of a difference in the distance modulus between $z$\,=\,2.8 and $z$\,=\,3.1, 
objects with same luminosities would have  apparent magnitudes 0.27\,mag fainter at $z$\,=\,3.1 than at $z$\,=\,2.8.
Therefore, we consider LAEs with $m_{NB}\leq 24.5$ at $z$\,=\,2.8 and those with $m_{NB}\leq 24.77$ at $z$\,=\,3.1, and then count their numbers in each narrowband.
Assuming that they all have the same volume as in NB475, the cumulative numbers in each narrowband at $z\sim$2.8 are 89, 50 and 25 for LAEs in NB466,
NB470 and NB475, respectively, while the numbers at $z\sim$3.1 are 49 (NB497 in SSA22-b) and 75 (NB501 in ECDFS). The bright LAE number, and hence the number density in NB466, is higher than that in the core SSA22 field by 81.6 \%, and by 18.7\% than that in ECDFS at $z$\,=\,3.1. It shows again that, NB466 field is a rare overdense region. Furthermore, there are 6 brightest LAEs with $m_{NB} \leq 22.75$ in NB466, including a luminous QSO.  Note that here we only compare the number of LAEs. In SSA22, several Lyman-alpha Blobs \citep[LABs,][]{Matsuda04} and AGNs \citep{Lehmer09} are reported, which are not accounted in the above comparison.

The projected distributions of these LAEs on the sky are shown in Figure \ref{spatial}, from which the overdensity of LAEs in the NB466 field is clearly shown.
To better visualize and quantify the distributions of LAEs, we create two-dimensional density maps in these fields. We apply a Gaussian kernel of FWHM = 5 arcmin (9.2 comoving Mpc) to the position maps of all LAE candidates, LAE candidates in each field, and spectroscopically confirmed and unconfirmed LAEs, respectively. The contours show the relative local density with respect to the density for all LAE candidates ($\rho_{avg}$). Obviously, the LAE distribution in the NB466 field is highly inhomogeneous. In particular, four high-density regions (PC1--4, within a radius of 5.5 comoving Mpc, see the bottom panel of Figure \ref{spatial})  contain 12, 9, 11 and 9 LAEs, respectively. One of the high-density regions is connected with a BAL-QSO at $z\sim$ 2.8 with density $\geq$ 3\,$\rho_{avg}$. Given that the observed average LAE surface density in the blank field is 0.055 arcmin$^{-2}$, the expected number of LAEs within a 3\arcmin\ radius circle in NB466 is $\sim$1.6$\pm$0.2 (after considering the overlap of NB466 and NB470). Hence, these four regions, PC1, PC2, PC3, and PC4, have overdensities of 6.6, 4.7, 6.0, and 4.7, respectively.

\subsection{Masses of the Protoclusters at $z\sim$ 2.8}
\label{sec:protocluster}

The overdensity region traced by high-redshift galaxies will eventually evolve into a bound, virialized system at $z$ = 0. The associated virialized mass (M$_{z=0}$) is: 
\begin{eqnarray}
M_{z=0} & = & (1+\delta_m)\langle \rho \rangle V, \nonumber\\
 & = & [1.37\times 10^{14} M_\odot](1+\delta_m)(V/(15 Mpc)^3),
\end{eqnarray}
where $\langle \rho \rangle$ is the average matter density of the universe (=[$3H_0^2/8\pi G]\Omega_m$), $\delta_m$ is matter overdensity, and $V$ is the volume enclosing the observed galaxy overdensity. We follow \citet{Steidel98,Steidel05} to measure the total mass overdensity $\delta_m$ from the observed galaxy overdensity $\delta_{g}$ from the equation 1+$b\delta_m$ = $C(1+\delta_g)$, where $C$ is the correction factor for correcting the redshift-space distortion and can be expressed as $C(\delta_m,z)$ = 1+$\Omega_m^{4/7}(z)[1-(1+\delta_m)^{1/3}]$ \citep{Steidel98,Steidel05}. We take LAE bias value of $b_{LAE} \approx$ 2 \citep[e.g.,][]{Gawiser07}. With $\delta_g\approx$  4.7--6.6, we find the parameter $C \approx$ 0.67--0.60 and $\delta_m\approx$ 1.41--1.79 ($\delta_m\approx$ [1.79, 1.41, 1.68, 1.41] for [PC1, PC2, PC3, PC4]). 

With the assumptions above and $\delta_g \approx$ 4.7--6.6, the total mass associated with each of the observed galaxy overdensities is 3.8 $\times$ 10$^{14}$ M$_\odot$, 3.3 $\times$ 10$^{14}$ M$_\odot$, 3.7 $\times$ 10$^{14}$ M$_\odot$, and 3.3 $\times$ 10$^{14}$ M$_\odot$ for the PC1, PC2, PC3, and PC4 protoclusters, respectively. The total volume enclosed in each protocluster is comparable to the cubic volume (15Mpc)$^3$ (i.e., $\pi(5.5\rm Mpc)^2\times 43\rm Mpc\times 0.83 \approx 15^3$ Mpc$^3$). Using cosmological simulations, \citet{Chiang13} found that an overdensity region with $\delta_g >$ 3.5 at $z\sim$ 3 in a (15Mpc)$^3$ volume have $>$ 80\% probability to evolve into a galaxy cluster (see table 4 of their paper). If evolving independently, each of the overdensity regions at $z\sim$ 2.8 will evolve into a Virgo-like cluster ($M$ = (3--10) $\times 10^{14}$ M$_\odot$).
Therefore these four overdensity regions are four protoclusters at $z\sim$ 2.8.
Because these protoclusters are connected together, we expect that the 4 protoclusters will eventually merge and form a massive cluster similar to the Coma cluster ($M$ $>$ 10$^{15}$ M$_\odot$).

 \begin{figure}
\includegraphics[angle=270,width=\linewidth]{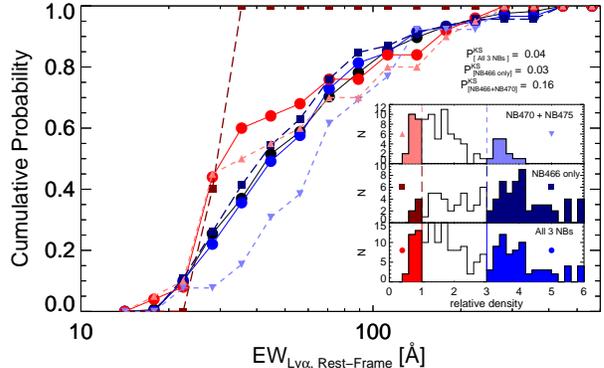}
\caption{The normalized cumulative \lya\ EW distributions of LAEs in different regions. The inset figure presents the relative LAE-density 
distributions in the whole field and the two separated fields (NB466 only and NB470+NB475), with red filled 
histograms marking the void regions, and blue filled histograms marking the overdense regions. $K-S$ test shows that the differences between 
EW distribution of LAEs in overdense regions and in void regions of NB466 and the whole field are $\gtrsim$ 2$\sigma$ (1 - $P_{KS}$ $\sim$ 96--97\%).}
\label{ksewtest}
\end{figure}

\subsection{\lya\ Luminosity Functions in the Overdense and Blank Fields}
\label{sec:lyalfdiff}

It is known that the luminosity functions should be different in overdense and blank fields. Performing the same \lya\ LF measurements in the overdense field and blank field would help us understand the effect of the clustering. Here we compare the luminosity function in the overdense field (NB466) and that in the blank filed (NB470+NB475).
We find the best-fit parameters of log$_{10}$($L^*$) = 42.84$\pm$0.13 and log$_{10}$($\Phi^*$) = -3.22$\pm$0.15 ($\chi^2/dof$ = 11.0/8) for the NB466 sample, and 
log$_{10}$($L^*$) = 42.49$\pm$0.10 and log$_{10}$($\Phi^*$) = -3.04$\pm$0.17 ($\chi^2/dof$ = 12.6/9) for the NB470+NB475 sample. Obviously, there are more bright LAEs in the overdense field (larger $L*$). 

The \lya\ photon density ($\propto$ $L^*\Phi^*$) in the overdense field is $\sim$50\% more than that in the blank field. The contours of the fitting parameters are plotted in Figure \ref{lyalflphi}. We find a difference of $\gtrsim 2\sigma$ between the \lya\ LFs in the overdense field and in the blank field. As would be expected, the \lya\ LF in the blank field at $z\sim$ 2.8 is located on the evolutionary path of the blank field \lya\ LFs from $z\sim$ 3.1 to $z\sim$ 2.1.

In the \lya\ LF fitting process, we exclude the 3 brightest LAEs (see Section \ref{sec:uvxdet} and Table \ref{xdet}), all of 
which are found in NB466 and detected in X-ray. The total contribution of the three LAEs to the \lya\ density in NB466 field 
is 1.3$\times$10$^{39}$ erg s$^{-1}$ Mpc$^{-3}$, which is about 22\% of the \lya\ density calculated from the \lya\ LF in 
NB466, and 32\% of that from NB470+NB475. It indicates that AGNs are not the major contributor of the \lya\ photons.

\subsection{The Environmental Effect on the Physical Properties of LAEs}
\label{sec:protoclusterphy}

\begin{figure}
\includegraphics[angle=270,width=\linewidth]{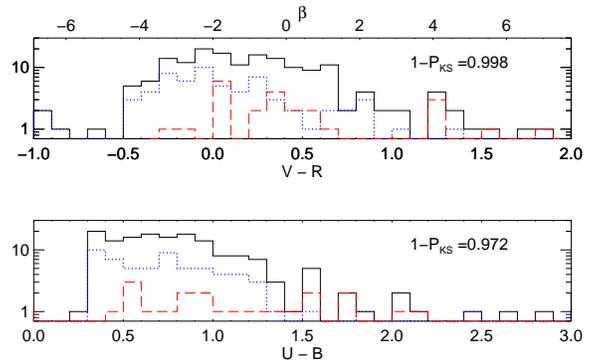} 
\caption{The color $U-B$ and $V-R$ distributions of all LAEs (black solid histograms),  LAEs in overdense regions (blue dotted histograms), and LAEs in void regions (red dashed histograms). 
$K-S$ test shows that the LAEs in overdense regions and void regions have different $U-B$ color distribution (1 - $P_{KS}$ = 97.2\%) and $V-R$ color distribution (1 - $P_{KS}$ = 99.8\%).
 The LAE subsamples in overdense and void regions are consistent with that in Figure \ref{ksewtest}.}
\label{kscolortest}
\end{figure}

We explore the \lya\ properties of LAEs in overdense ($>$ 3\,$p_{avg}$) and void ($<$ 1\,$p_{avg}$) regions. Through 
$K$-$S$ test, the \lya\ flux distribution is similar (1 - $P_{KS}$ = 25\%), while the \lya\ EW distribution in overdense 
and void regions is different (1 - $P_{KS}$ $\sim$ 96\%, see Fig. \ref{ksewtest}). This difference is slightly more significant 
in NB466 field (1 - $P_{KS}$ $\sim$ 97\%). In the spatial distribution plot, LAEs with large EWs are 
mostly located within the overdense lines within $\sim$ 3\,$p_{avg}$ contour-level (Fig. \ref{spatial}) around the clustering regions. 
\citet{Yamada12} also reported similar difference between the EW distributions of LAEs in overdense regions and those 
in void regions of SSA 22 field. The case of similar \lya\ flux distribution but lower \lya\ EWs in void regions than overdense 
regions implies that  \lya\ photons of LAEs in overdense regions are more easily to escape. Further explorations on the SFR 
and $UBVR$ magnitudes distribution of LAEs in overdense and void regions do not show any significant difference. However, 
on the distributions of the $U$-$B$ and $V$-$R$ colors we find that LAEs in overdense regions are bluer than that of LAEs in 
void regions with a significant level of 1 - $P_{KS}$ =  97.2\% and 99.8\%  (see Fig. \ref{kscolortest}). Considering the relation 
between UV slope $\beta$ and color $V$-$R$, LAEs in overdense regions have much steeper UV slope $\beta$ than LAEs in 
void regions. Through the population synthesis model, galaxies with steeper UV slope are younger and have more massive stars (excluding AGN contribution). 
LAEs in overdense regions also have smaller $U$-$B$ values, which implies a larger UV escape fraction of LAEs in overdense 
regions. The explanations of small $U$-$B$ values include less dust and more hard UV photons, which are common in young galaxies. 

The significant differences on \lya\ EW distributions and broad band color distributions of LAEs in overdense and void regions 
imply that LAEs in overdense regions (mainly in protoclusters) are younger and possible less dusty than LAEs in voids at 
$z$\,=\,2.8. The overdensity itself may cause these characteristics, i.e., the gravitational potential of the overdense region 
would trap the ISM gas and increase the inflow. The high merger rates in clusters would also help to generate more UV photons 
via intense star-burst.

The similar \lya\ flux distribution in the overdense and blank fields
indicates that there may be no environment dependence in star formation
activities at $z>$ 2, which was shown in previous studies \citep[e.g.,][]{Hayashi12, Cooke14}. 
The younger age in the overdense region is a
sign of having more newly formed stars in protoclusters. In the meanwhile,
there are likely more dusty starburst galaxies in the protoclusters of
\citet{Hayashi12} and \citet{Cooke14}. Given that protoclusters in
their work are selected through Ha imaging which is less dust sensitive,
our result is consistent with theirs that protocluster environment is
favorable to the new star formation. This is in contrast to the
fact in the local universe that old red galaxies tend to occupy dense
regions while blue star-forming galaxies are more likely to be found in
blank fields \citep[e.g.,][]{Dressler80}. However, it is still uncertain in our
studies if the star formation properties are the same for galaxies with the
same stellar mass in different environments. Further SED analysis is needed
to examine the SFR-stellar mass relations and the respective mass functions
both in the overdense region and in the blank field, so that to quantify
the environmental effects at $z\sim$ 2.8 (Zheng et al. in Prep).


\section{Summary and Conclusions}
\label{sec:conclusion}

Using three narrowband filters (NB466, NB470, and NB475), we have carried out a narrowband survey for Lyman Alpha Emitter galaxies at $z\sim$ 2.8-2.9 in the ECDFS field. Then we have performed a spectroscopic follow-up observation of the LAE candidates selected from the narrowband survey.
A large scale structure at $z\sim$2.8 is confirmed with the overdense distribution of LAEs in NB466, which is indicated by a peak in 
the redshift distribution of galaxies in ECDFS in previous work. 
We have found that, this large scale structure consists of 4 protoclusters, whose overdensities within a radius of 5 Mpc (equivalent comoving volume $15^3$Mpc$^3$) relative to the blank field (NB470+NB475) are in the range of 4.7 - 6.6, and the overdense structure is expected to evolve into a Coma-like cluster.

In the meanwhile, we have investigated the various physical properties of LAEs at these redshifts, including star formation rates and luminosity functions. The environmental effects are also studied through the comparison between LAE properties in the overdense field field (NB466) and in the blank field (NB470+NB475).

The main points of this paper are summarized as below: 
\begin{enumerate}
\item  \textbf{LAE selection at $z\sim$ 2.8:} In the narrowband survey, we use a color cut of 
($U$-$B$ $\geq$ 0.3) $\|$ ($U$-$B$ $<$ 0.3 \& $B$/$\sigma_B$ $<$ 2)  to select LAE candidates at $z\sim$ 2.8, which 
has a spectroscopic success fraction of $\sim$ 74\%, a contamination fraction of $\sim$ 7--26\%, and a complete fraction 
of $\sim$ 86\%.  After excluding low-z interlopers detected by $GALEX$ and other spectroscopic surveys, 
we obtain a complete sample of 179 LAEs (96 from NB466, 55 from NB470, and 28 from NB475) at $z\sim$ 2.8--2.9 with 
F$(Ly\alpha)$ $\geq$ 2.9$\times$ 10$^{-17}$ erg\,cm$^{-2}$\,s$^{-1}$ and rest-frame EW$(Ly\alpha)$ $\geq$ 20\AA. 

\item  \textbf{Overdensity in NB466:} The overdensity in NB466 is confirmed by the number excess of LAEs.
Comparable to the density excess of $\delta_{LAE}$ = 1.13$\pm$0.01 for SSA22 Sb1, the NB466 has 
$\delta_{LAE}$ = 1.34$\pm$0.24 (compared with the two other narrowband fields).   This is $>$5 times the standard 
deviation for LAEs estimated from the fluctuations of dark matter and LAE bias value of 2. We also compare our 
surveys to the panoramic surveys of LAEs at $z\sim$ 3.1 \citep{Yamada12}. There are more bright LAEs in NB466 
compared with other fields. 

\item \textbf{The large-scale structure and protoclusters at $z\sim$ 2.8:} The large-scale structure at $z\sim$ 2.8 is 
clearly shown in the spatial distribution of LAEs. It composes 4 overdense regions each within a radius of 3\,$\arcmin$. 
The observed LAE numbers in the four protoclusters are in the range from 9 to 12, implying galaxy overdensities of 
$\delta_g\sim$ 4.7--6.6. The four protoclusters are connected together, and will evolve into a Coma-like 
cluster (M $>$ 10$^{15} $M$_\odot$) at present day.

\item \textbf{Spectroscopic observations:} In the spectroscopic followup,  41 of 55 targets from the complete sample 
have been confirmed. We did not find lines other than the \lya\ in the individual spectra.  
The stacked spectrum of confirmed LAEs has a steep slope of $\beta$ $\sim$ -2.5, and a 4.5-$\sigma$ detection of 
C\,{\sc iii}] line, with rest-frame EW of 10\AA\ and flux ratio of 1/8 to \lya. There is an offset of $\sim$ 
300$\pm$147 km s$^{-1}$ between the C\,{\sc iii}] and \lya\ lines in the stacked spectrum, consistent with 
the \lya\ offsets of LAEs reported at $z$\,$\sim$\,2--3.

\item \textbf{SFRs of LAEs at $z\sim$ 2.8:} The SFRs of LAEs at $z\sim$ 2.8 are checked via multiple indicators, 
including SFRs estimated from the \lya\ line (SFR$_{Ly\alpha}$), the rest-frame UV radiation from V-band 
(SFR$_{UV_V}$), and X-ray (SFR$_X$). On average, we get SFR$_{Ly\alpha}$ of $\sim$4 M$_\odot/yr$ in each 
narrowband, but higher SFR$_{UV_V}$ of $\sim$9-11 M$_\odot/yr$. The average X-ray signal of LAEs without 
X-ray detection is $<$ 1.5\,$\sigma$, implying a 1-$\sigma$ upper limit of SFR$_X$ $<$ 16 M$_\odot/yr$. With 
these values we can estimate that the \lya\ escape fraction is in the range 25\% $\lesssim f_{ESC}^{Ly\alpha} 
\lesssim$ 40\% (SFR$_{Ly\alpha}$/SFR$_X$ $\lesssim f_{ESC}^{Ly\alpha} \lesssim$ SFR$_{Ly\alpha}$/SFR$_{UV}$), 
and the escape fraction of UV continuum photons is $f_{ESC}^{UV,cont.}$ $\gtrsim$ 62\% ($f_{ESC}^{UV,cont.}$ 
$\gtrsim$ SFR$_{UV}$/SFR$_X$) for LAEs at $z\sim2.8$. The X-ray constraint on the  \lya\ escape fraction at 
$z\sim$ 2.8 is consistent with the Far-infrared constraint by \citet{Wardlow14} and \citet{Kusakabe15}. 

\item \textbf{\lya\ LF at $z\sim$ 2.8:} The \lya\ luminosity functions at $z\sim$ 2.8--2.9 are located within the 
values between $z=$ 2.1 and 3.1, which is consistent with the general galaxy evolution between 2 $<$ $z$ $\lesssim$ 3. 
The \lya\ LF at $z\sim$ 2.8--2.9 from the complete sample is fitted with a Schechter function with parameters of 
log$_{10}$($L^*$) = 42.73$\pm$0.08, log$_{10}$($\Phi^*$) = -3.21$\pm$0.11 and a fixed $\alpha$ = -1.5.  
When including faint luminosity bins, we obtain the faint-end slope of $\alpha$ = -2.1$\pm$0.3.

\item \textbf{SFRDs from \lya\ and UV:} The \lya\ and UV-based cosmic SFRDs are compared. By integrating 
the  \lya\ luminosity function over luminosity and case-B recombination, we obtain the cosmic SFRD from \lya. 
For LAE surveys at $z$ = [2.1, 2.8, 3.1, 3.7], the cosmic SFRDs from \lya\ are log$_{10}$($\Psi_{Ly\alpha}(z)$) = 
[-2.49$\pm$0.11, -2.37$\pm$0.10, -2.25$\pm$0.06, -2.33$\pm$0.12], and peak around $z=3.1$ (1--2$\sigma$). 
The ratio between the \lya\ and UV-based SFRDs is $\lesssim$ 0.03 at $z <$ 2, while slightly increasing from 0.1 
to $z\sim$ 4.5, then showing a large scatter started from $z\sim$ 5.7. If the observed \lya\ SFRDs are constant over the 
redshift range 2.8 $\lesssim z\lesssim$ 5.7, the  ratio of \lya\ photons from LAEs to UV photons from high-redshift 
galaxies would increase as a function of redshift. This is consistent with the increasing fraction of galaxies with 
strong \lya\ emissions from $z\sim$ 3 to $z\sim$ 6 reported by \citet{Stark10}.

\item \textbf{\lya\ LFs in overdense and void fields:} The \lya\ Luminosity Functions in the overdense field (NB466) 
and the blank field (NB470+NB475) show a $\gtrsim$ 2$\sigma$ difference. The best-fit parameters are 
log$_{10}$($L^*$) = 42.84$\pm$0.13 and log$_{10}$($\Phi^*$) = -3.22$\pm$0.15 ($\chi^2/dof$ = 11.0/8) for 
LAEs in the overdense field, and log$_{10}$($L^*$) = 42.49$\pm$0.10 and log$_{10}$($\Phi^*$) = -3.04$\pm$0.17 
($\chi^2/dof$ = 12.6/9) for LAEs in the blank field. There are more bright LAEs in the overdense field, and 
the \lya\ photon density ($\propto$ $L^*\Phi^*$) in the overdense field is $\sim$50\% more than that in the 
blank field for LAEs at $z\sim$ 2.8.

\item \textbf{Environmental effects on LAE properties:} We explore the physical properties of LAEs in different 
environments. The distributions of \lya\ flux and broadband magnitudes of LAEs in overdense and void regions are nearly same.  However, a 
difference of $\sim$2--3 $\sigma$ is shown in  the distributions of the EWs, the $U$-$B$ color, and the $V$-$R$ 
color of LAEs in overdense and void-regions. LAEs in overdense regions are younger and possible less dusty than 
that in void regions, and the clustering itself may cause these characteristics.

\end{enumerate}


\acknowledgments

We would like to thank the anonymous referee for the constructive comments that led us to substantially improve the quality of this paper. Z.Y.Z gratefully acknowledges support from the Chinese Academy of Sciences (CAS) through a CAS-CONICYT Postdoctoral Fellowship administered by the CAS South America Center for Astronomy (CASSACA) in Santiago, Chile. 
J.X.W. thanks support from NSFC 11421303. C.Y.J. is supported by Shanghai Natural Science Foundation (15ZR1446600).
This work was developed during the stay of Z.Y.Z. as SESE Exploration postdoctoral fellow at the Arizona State University. Z.Y.Z. would like to thank Roderik A. Overzier and Yi-kuan Chiang for helpful suggestions and comments. We thank the staff of Cerro Tololo Inter-American Observatory and Las Campanas Observatory for their expert assistance throughout this project. 

Based on observations at Cerro Tololo Inter-American Observatory, National Optical Astronomy Observatory (NOAO Prop. ID: 2011B-0569; PI: Z. Zheng), which is operated by the Association of Universities for Research in Astronomy (AURA) under a cooperative agreement with the National Science Foundation. 

{\it Facilities:} \facility{Blanco(MOSAIC-2)}, \facility{Magellan:Baade(IMACS)}

\appendix
\label{appendix}

We derive the calculations for the pure emission line flux and EW from narrowband and broadband magnitudes in this Appendix section. We also present how the filter profile and observational uncertainties would change the estimations of the line flux and EW, and thus bias the distributions of \lya\ LF and EW calculated from the observed LAE sample. \\

\section{A-1: Calculation of Line Flux and EW}

Assuming that the emission-line candidates are LAEs at $z$\,$\sim$\,2.8--2.9, their \lya\ fluxes $F_{Ly\alpha}$ and \lya\ EWs can be calculated following the formulae in \citet{Zheng14}. 
The steps are summarized as below. A fake LAE spectrum with \lya\ emission line and UV continuum ($f_\lambda \propto \lambda^{\beta}$) modified by IGM absorption $C_{\textsc{IGM}}(\lambda, z)$ \citep{Madau95} is selected:
\begin{eqnarray}
f_{\textsc{mod},\lambda} & = & f_{\textsc{con},\lambda_{Ly\alpha}} [ \delta(\lambda_{Ly\alpha}) \textsc{EW}_{obs} +  ({\lambda}/{\lambda_{Ly\alpha}}) ^{\beta} \textsc{C}_{\textsc{IGM}}(z,\lambda)] , \nonumber \\ 
 & = & f_{\textsc{line},\lambda} + f_{\textsc{con},\lambda},
\end{eqnarray}
here the \lya\ wavelength $\lambda_{Ly\alpha} = (1+z)\times1215.67$\AA. Then the narrow-band and broad-band flux densities can be modeled as: 
\begin{eqnarray}
f_{\nu, NB} & = &  \Big[\int f_{\textsc{line},\lambda} f_{NB,\lambda} d\lambda + \int f_{\textsc{con},\lambda} f_{NB,\lambda} d\lambda \Big] \times \frac{\lambda_{NB}^2}{c W_{NB}}, \nonumber\\
                  & = & \Big[ a_N \times F_{Ly\alpha} + b_N \times \frac{F_{Ly\alpha}}{EW_{obs}} \times W_{NB} \Big]\times \frac{\lambda_{NB}^2}{c W_{NB}}, \\
f_{\nu, BB} & = &  \Big[\int f_{\textsc{line},\lambda} f_{BB,\lambda} d\lambda + \int f_{\textsc{con},\lambda} f_{BB,\lambda} d\lambda \Big] \times \frac{\lambda_{BB}^2}{c W_{BB}}, \nonumber\\
                  & = & \Big[ a_B \times F_{Ly\alpha} + b_B \times \frac{F_{Ly\alpha}}{EW_{obs}} \times W_{BB} \Big]\times \frac{\lambda_{BB}^2}{c W_{BB}}.                 
\end{eqnarray}
Here $f_\nu(BB)$, $ f_\nu(NB)$, $\lambda_{BB}$, $W_{BB}$ and $W_{NB}$ are the $B$ band flux density, the $NB$ band flux density, 
the $B$ band central wavelength, the $B$ band bandwidth and the $NB$ band bandwidth , respectively.  
We calibrate narrow-band to B-band, $\lambda_{NB}$ = $\lambda_{BB}$. The coefficients 
$b_B$, $b_N$, $a_N$ and $a_B$ should be 1 under the approximation of a top-hat filter and no IGM absorption.  When considering the filter profile and IGM absorption, the 
coefficients $b_B$ = $\int f_{con,\lambda} f_{BB}d\lambda/[\int (f_{con,\,\lambda\,=\,\lambda_{Ly\alpha}}  f_{BB}) d\lambda]$ and $b_N$ = $\int f_{con,\lambda} f_{NB}d\lambda$/[$\int f_{con,\,\lambda\,=\,\lambda_{Ly\alpha}} f_{NB} d\lambda]$ account for IGM absorption \citep{Madau95} of continuum in B band and narrowband, respectively. The coefficients 
$a_N$ = $\int f_{line,\lambda} f_{NB}d\lambda$/[$\int f_{line,\lambda} d\lambda\times$max($f_{NB})]$  and $a_B$ = $\int f_{line,\lambda} f_{BB}d\lambda/[\int f_{line,\lambda} d\lambda\times$max($f_{BB})]$ correct the effect when the narrowband and B band filters are in the not top-hat shapes. Solving the equations above, we have:
\begin{eqnarray}
F_{Ly\alpha} & = & \frac{(b_B f_{\nu, NB} - b_N  f_{\nu, B}) W_{NB} W_{BB} c}{(a_N  b_B W_{BB} - a_R b_N  W_{NB}) \lambda_{BB}^2} \\
 EW_{obs} 	&  =   &  \frac{ (b_B f_{\nu, NB} -b_N f_{\nu, B}) W_{BB}  W_{NB}}{(a_N f_{\nu,B} W_{BB} -a_B f_{\nu,NB} W_{NB})} , 
\end{eqnarray}
 We choose $\beta$ = -2 in the calculations,  and get typical values of $a_N$ = [1.00,0.95,0.97], $a_B$ = [0.96,0.98,1.00], $b_N$ = [0.87,0.82, 0.91] and $b_B$ = [0.88, 0.88, 0.88]
 for \lya\  flux and EW calculations of LAEs in NB466, NB470, and NB475, respectively. The change of $\beta$ by $\pm$0.5 would introduce $<$5\% errors of \lya\ flux and EW calculations.  
 Note that here we assume the \lya\ line is located at the center of the corresponding narrowband filter.

\section{A-2: The redshift (filter profile) dependence on EW and Flux estimations}

\begin{figure*}
\includegraphics[angle=0,width=\linewidth]{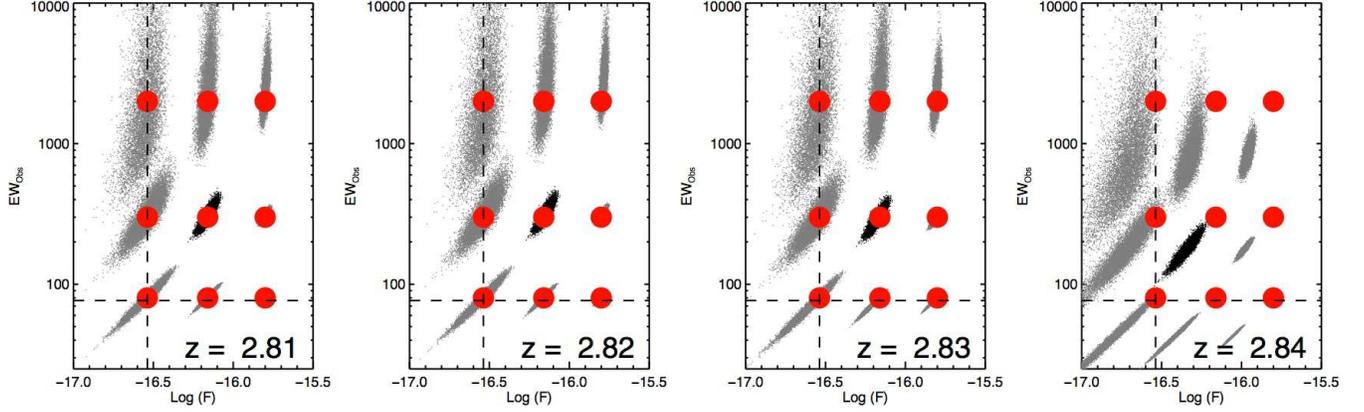} 
\caption{(Appendix-1) Intrinsic flux and EW vs. simulated observational estimations of fluxes and EWs at different redshifts for LAEs in NB466. Here different redshifts are corresponding to different locations on the profile of the narrowband filter NB466. For each input line flux and EW (red filled circles), we mock 10,000 LAE spectra at the corresponding redshifts, and then convolve with narrowband and broadband filters, add observational errors to generate observed narrowband and broadband magnitudes, and finally estimate their line fluxes and EWs with Eq. A4 and A5 (black or grey dots). 
These plots demonstrate how observational errors and the line locations on the profile of the narrowband filter would change the estimations of the intrinsic EWs and fluxes.  The horizontal and vertical dashed lines mark the ranges of our complete LAE sample (c). Obviously, the observational estimations underestimate the input line fluxes and EWs at the edge of the narrowband filter (e.g., z=2.84).    }
\label{ewdist}
\end{figure*}

The above Ly$\alpha$ fluxes and EWs are estimated under the assumption that the \lya\ line is located at the central wavelength of the narrowband filter.  
When the \lya\ line shifts out of the center wavelength, the above calculations underestimate both the \lya\ EWs and the \lya\ line fluxes. The underestimations are significant when the line is shifted to the edge of the narrowband filter profile (i.e., for z=2.84 LAEs in NB466). The underestimations lead to incomplete sample selection at the tail of the filter profile (see the right panels of Fig. \ref{ewdist} for z=2.84 LAEs in NB466).

We mock this process by generating fake LAE spectra with input line flux and EW values at different redshifts, then convolving with broadband and narrowband profiles, adding observational errors, and finally using Equations A4 and A5 to estimate the observed line flux and EW values. For each input line flux and EW pair, we mock 10,000 random observational errors on narrowband and broadband magnitudes. The distributions of output line flux and EW pairs at different redshifts are presented in Figure \ref{ewdist2}. Obviously, when the input line is located at the edge of the narrowband filter profile, Equations A4 and A5 would systematically underestimate the input line flux and the EW.  More interestingly, observational errors would bias the estimations of the line flux and the EW largely when the input EW is large. Therefore the selection of faint LAEs and LAEs with EW$_r$ $\sim$ 20\AA\ is incomplete when including observational errors. 

\section{A-3: Recovering the \lya\ LF and EW distributions}

 In order to obtain the true distributions of Ly$\alpha$ line luminosities and EWs from observations, we follow \cite{Zheng14} to use a Monte Carlo approach to generate a mock LAE catalog with prior known \lya\ LF and EW distributions.  We then add observational errors and apply the LAE selection processes described in Section 2.1. Finally we compare the input distributions of \lya\ fluxes and EWs to the mock output distributions of \lya\ fluxes and EWs. In this step the redshift dependence (Section A-2) is also taken into account. 

\begin{figure*}
\includegraphics[angle=0,width=\linewidth]{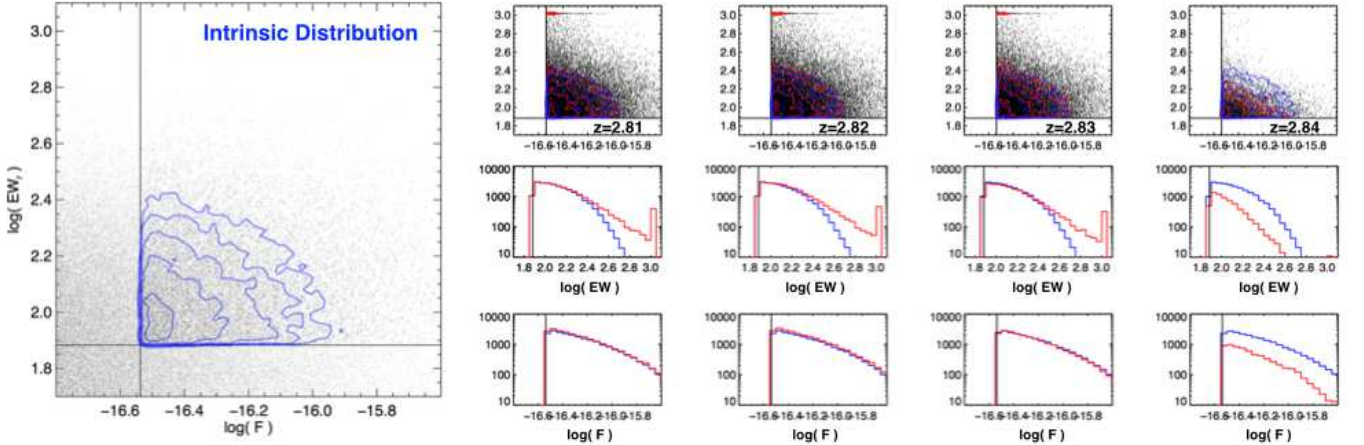} 
\caption{(Appendix-2) Left panel: the input distributions of line fluxes and EWs following the distributions of \lya\ LF and EWs at $z\sim$ 3 \citep[e.g., from][Log(L$*$) = 42.75 and EW$_{r,0}$ = 65 \AA]{Ciardullo12}. Right top panels: the output distributions of line fluxes and EWs for LAEs at z=2.81, 2.82, 2.83, and 2.84. Right middle panels: the output EW distributions (red histogram) compared with the input EW distributions (blue histogram) for LAEs at the corresponding redshifts. Right bottom panels: the output line flux distributions (red histogram) compared with the input line flux distributions (blue histogram) for LAEs at the corresponding redshifts.
}
\label{ewdist2}
\end{figure*}

We generate a mock LAE catalog in NB466 to estimate the selection effect on the redshift dependence.
The mock LAE catalog is generated following the \lya\ LF and EW distributions at $z\sim$ 3 \citep[e.g., from][Log(L$*$) = 42.75 and EW$_{r,0}$ = 65 \AA]{Ciardullo12}.  There are 139,803 mock LAEs with intrinsic luminosity in the range of 42 $\leq$ Log(L) $\leq$ 43.5 and rest-frame EW in the range of EW$_rest$ $\geq$ 5\AA (see left panel of Figure \ref{ewdist2}), and $\sim$54k mock LAEs with intrinsic flux and EW in the complete LAE sample ranges ($F$ $\geq$ 2.9$\times10^{-17}$ erg\,cm$^{-2}$\,s$^{-1}$ and EW $\geq$ 20\AA).  Then we simulate the observational uncertainties and select the output sample following the methods introduced above. The two dimensional distribution of input and output line fluxes and EWs are presented in Figure \ref{ewdist2} (here we take NB466 as an example). Our NB466 filter is a nearly 'top-hat' filter, and thus LAEs at z=2.81, 2.82 and 2.83 are well recovered. At the edge of the filter, i.e., LAEs at z=2.84, a large sample of LAEs are missed because of the underestimations. The fractions of complete LAE sample recovery are [27.7\%, 102.7\%, 103.9\%, 92.7\%, 46.6\%] for AEs at $z\sim$ [2.80, 2.81, 2.82, 2.83, 2.84] in NB466 narrowband imaging. 

The input and output distributions of line fluxes and EWs are plotted in the bottom and central panels of Figure \ref{ewdist2}. The observational uncertainties and selection processes do not significantly change the shape of flux distributions. However, they significantly change the EW distributions. Observational uncertainties systematically boost EW values, therefore in this work we ignore the EW distribution of the whole sample.

\clearpage

\begin{deluxetable*}{lcccccccccccc}
\tablefontsize{\scriptsize}
\tablecaption{The catalog of spectroscopically confirmed emission-line galaxies at $z\sim$ 2.8 in the ECDFS field.\label{tbl-1}}
\tablewidth{0.99\textwidth}
\tablehead{
\colhead{Obj.} & \colhead{RA} & \colhead{DEC} & \colhead{z} & \colhead{Q$^a$} &
\colhead{F(\lya)$^b$} & \colhead{EW$_R$(\lya)} & \colhead{NB} &
\colhead{$U^c$} & \colhead{$B$} &
\colhead{$V$} & \colhead{$R^c$}
}
\startdata
N466\_043 &   53.1853 &  -27.9102 &  2.807 & 3 &     3.54$\pm$0.54 &   9000$^{+      0.0} _{-  7650}$ &    24.76$\pm$0.16 &    28.80$\pm$0.68$^{ V}$ &    28.57$\pm$0.99 &    28.99$\pm$2.98 &    26.90$\pm$0.15$^{V}$  \\ 
N466\_047 &   53.1812 &  -27.9025 &  2.810 & 3 &    10.83$\pm$0.90 &     41.3$^{+      4.9} _{-     4.9}$ &    23.24$\pm$0.07 &    25.02$\pm$0.03$^{ V}$ &    24.66$\pm$0.04 &    24.57$\pm$0.08 &    24.56$\pm$0.03$^{V}$  \\ 
N466\_056 &   52.9535 &  -27.8929 &  2.822 & 3 &    10.48$\pm$1.04 &     89.3$^{+     22.2} _{-    18.3}$ &    23.42$\pm$0.09 &    26.23$\pm$0.23$^{ V}$ &    25.36$\pm$0.11 &    24.95$\pm$0.15 &    24.79$\pm$0.14$^{M}$  \\ 
N466\_065 &   53.3457 &  -27.8763 &  2.807 & 3 &    12.32$\pm$0.99 &     53.4$^{+      7.3} _{-     6.9}$ &    23.16$\pm$0.07 &    25.77$\pm$0.43$^{ M}$ &    24.75$\pm$0.06 &    24.61$\pm$0.11 &    24.49$\pm$0.10$^{M}$ \\ 
N466\_067 &   53.0976 &  -27.8698 &  2.804 & 3 &     4.57$\pm$0.69 &   9000$^{+      0.0} _{-  8674}$ &    24.46$\pm$0.15 &    27.11$\pm$0.20$^{ V}$ &    27.70$\pm$0.67 &    27.78$\pm$1.55 &    26.95$\pm$0.20$^{V}$  \\ 
N466\_069 &   53.0265 &  -27.8680 &  2.814 & 3 &    13.50$\pm$0.92 &     91.7$^{+     14.8} _{-    13.4}$ &    23.14$\pm$0.06 &    25.52$\pm$0.06$^{ V}$ &    25.11$\pm$0.08 &    24.39$\pm$0.08 &    24.86$\pm$0.07$^{V}$  \\ 
N466\_072 &   53.0461 &  -27.8637 &  2.818 & 3 &     2.89$\pm$0.40 &     42.5$^{+      9.4} _{-     8.6}$ &    24.68$\pm$0.11 &    26.62$\pm$0.08$^{ V}$ &    26.11$\pm$0.09 &    25.78$\pm$0.13 &    25.93$\pm$0.08$^{V}$  \\ 
N466\_078 &   53.2848 &  -27.8519 &  2.810 & 3 &     5.46$\pm$0.47 &     34.6$^{+      4.4} _{-     4.3}$ &    23.93$\pm$0.06 &    25.94$\pm$0.05$^{ V}$ &    25.23$\pm$0.05 &    25.07$\pm$0.08 &    25.21$\pm$0.03$^{V}$  \\ 
N466\_082 &   53.0511 &  -27.8483 &  2.828 & 3 &     4.68$\pm$0.61 &     57.7$^{+     14.5} _{-    12.1}$ &    24.22$\pm$0.11 &    26.39$\pm$0.08$^{ V}$ &    25.87$\pm$0.11 &    25.77$\pm$0.20 &    25.82$\pm$0.08$^{V}$  \\ 
N466\_084 &   53.1074 &  -27.8441 &  2.828 & 2 &     3.41$\pm$0.53 &     76.6$^{+     33.3} _{-    23.3}$ &    24.61$\pm$0.13 &    26.97$\pm$0.16$^{ V}$ &    26.44$\pm$0.19 &    25.80$\pm$0.22 &    25.90$\pm$0.07$^{V}$  \\ 
N466\_086 &   53.0573 &  -27.8408 &  2.826 & P &     4.06$\pm$0.77 &     29.2$^{+      7.8} _{-     7.1}$ &    24.20$\pm$0.13 &    25.99$\pm$0.07$^{ V}$ &    25.38$\pm$0.09 &    24.93$\pm$0.12 &    25.13$\pm$0.05$^{V}$  \\ 
N466\_096 &   53.0863 &  -27.8187 &  2.813 & P &     7.13$\pm$0.91 &     42.0$^{+      8.4} _{-     7.8}$ &    23.69$\pm$0.10 &    25.44$\pm$0.05$^{ V}$ &    25.12$\pm$0.08 &    24.66$\pm$0.10 &    24.74$\pm$0.04$^{V}$  \\ 
N466\_105 &   53.0516 &  -27.8046 &  2.817 & 3 &    17.30$\pm$0.87 &     48.8$^{+      3.9} _{-     3.9}$ &    22.77$\pm$0.04 &    25.14$\pm$0.03$^{ V}$ &    24.31$\pm$0.03 &    23.70$\pm$0.04 &    23.85$\pm$0.01$^{V}$  \\ 
N466\_109$^d$ &   53.0393 &  -27.8019 &  2.817 & 4 &   171.29$\pm$0.94 &     30.7$^{+      0.2} _{-     0.2}$ &    20.16$\pm$0.00 &    23.25$\pm$0.01$^{ V}$ &    21.39$\pm$0.00 &    20.83$\pm$0.00 &    20.90$\pm$0.00$^{V}$ \\
N466\_112 &   53.2221 &  -27.7945 &  2.832 & -1 &     3.60$\pm$0.42 &     71.5$^{+     16.7} _{-    14.0}$ &    24.54$\pm$0.10 &    27.31$\pm$0.12$^{ V}$ &    26.34$\pm$0.10 &    26.20$\pm$0.17 &    26.40$\pm$0.09$^{V}$  \\ 
N466\_114 &   52.9101 &  -27.7933 &  2.818 & 3 &     8.01$\pm$0.58 &    144.0$^{+     35.8} _{-    29.5}$ &    23.76$\pm$0.07 &    27.28$\pm$1.10$^{ M}$ &    26.00$\pm$0.11 &    26.07$\pm$0.24 &    26.10$\pm$0.26$^{M}$  \\ 
N466\_122 &   53.3456 &  -27.7852 &  2.805 & 3 &    11.61$\pm$0.78 &    166.1$^{+     47.8} _{-    35.6}$ &    23.37$\pm$0.06 &    26.33$\pm$0.56$^{ M}$ &    25.70$\pm$0.12 &    25.80$\pm$0.26 &    25.88$\pm$0.11$^{V}$  \\ 
N466\_123 &   52.9545 &  -27.7847 &  2.832 & 2 &     3.40$\pm$0.62 &     72.4$^{+     31.3} _{-    22.0}$ &    24.61$\pm$0.16 &    26.71$\pm$0.12$^{ V}$ &    26.40$\pm$0.17 &    26.21$\pm$0.29 &    26.27$\pm$0.16$^{V}$ \\ 
N466\_126 &   53.3058 &  -27.7827 &  2.810 & 2 &    10.96$\pm$0.79 &     27.9$^{+      2.6} _{-     2.6}$ &    23.11$\pm$0.05 &    25.02$\pm$0.05$^{ V}$ &    24.27$\pm$0.03 &    23.35$\pm$0.03 &    23.09$\pm$0.01$^{V}$  \\ 
N466\_128 &   52.9575 &  -27.7805 &  2.830 & 3 &     3.06$\pm$0.75 &     41.6$^{+     16.6} _{-    13.6}$ &    24.61$\pm$0.19 &    26.79$\pm$0.15$^{ V}$ &    26.01$\pm$0.14 &    25.21$\pm$0.14 &    26.11$\pm$0.19$^{V}$  \\ 
N466\_137 &   53.1381 &  -27.7721 &  2.812 & 3 &     4.75$\pm$0.49 &     63.5$^{+     12.8} _{-    11.0}$ &    24.22$\pm$0.09 &    26.67$\pm$0.07$^{ V}$ &    25.94$\pm$0.09 &    25.45$\pm$0.11 &    25.70$\pm$0.06$^{V}$  \\ 
N466\_162 &   52.9617 &  -27.7274 &  2.825 & 3 &     3.27$\pm$0.40 &    122.7$^{+     50.0} _{-    35.2}$ &    24.72$\pm$0.11 &    27.84$\pm$0.19$^{ V}$ &    26.85$\pm$0.17 &    26.28$\pm$0.20 &    25.96$\pm$0.16$^{M}$  \\ 
N466\_164 &   53.2124 &  -27.7273 &  2.822 & 3 &     5.07$\pm$0.58 &     39.6$^{+      6.4} _{-     6.1}$ &    24.05$\pm$0.09 &    26.14$\pm$0.06$^{ V}$ &    25.44$\pm$0.06 &    25.03$\pm$0.08 &    25.29$\pm$0.05$^{V}$ \\ 
N466\_178 &   53.2066 &  -27.7032 &  2.825 & 3 &    17.36$\pm$0.99 &     70.8$^{+      7.8} _{-     7.4}$ &    22.83$\pm$0.05 &    25.27$\pm$0.05$^{ V}$ &    24.63$\pm$0.05 &    24.40$\pm$0.09 &    24.17$\pm$0.03$^{V}$ \\ 
N466\_186 &   52.9165 &  -27.6877 &  2.812 & 3 &     4.29$\pm$0.76 &     32.4$^{+      8.5} _{-     7.8}$ &    24.17$\pm$0.13 &    26.52$\pm$0.24$^{ V}$ &    25.42$\pm$0.10 &    24.95$\pm$0.13 &    25.02$\pm$0.16$^{M}$  \\ 
N466\_210 &   53.0127 &  -27.6004 &  2.810 & P &     9.75$\pm$1.35 &     44.7$^{+     10.5} _{-     9.3}$ &    23.37$\pm$0.11 &    25.19$\pm$0.11$^{ V}$ &    24.84$\pm$0.09 &    23.93$\pm$0.09 &    23.69$\pm$0.07$^{M}$  \\ 
N470\_035 &   53.0136 &  -27.9032 &  2.869 & 3 &     9.70$\pm$1.02 &     27.0$^{+      3.8} _{-     3.7}$ &    23.24$\pm$0.07 &    24.68$\pm$0.03$^{ V}$ &    24.37$\pm$0.05 &    23.74$\pm$0.05 &    24.06$\pm$0.05$^{V}$  \\ 
N470\_041 &   53.0107 &  -27.8828 &  2.855 & 3 &     5.20$\pm$0.84 &     33.8$^{+      8.1} _{-     7.4}$ &    23.98$\pm$0.12 &    25.57$\pm$0.07$^{ V}$ &    25.26$\pm$0.09 &    24.61$\pm$0.10 &    24.75$\pm$0.07$^{V}$  \\ 
N470\_047 &   53.1986 &  -27.8615 &  2.838 & 3 &     5.29$\pm$0.82 &     89.9$^{+     39.0} _{-    26.5}$ &    24.15$\pm$0.14 &    26.98$\pm$0.22$^{ V}$ &    26.10$\pm$0.18 &    25.82$\pm$0.27 &    25.31$\pm$0.06$^{V}$  \\ 
N470\_050 &   53.1924 &  -27.8443 &  2.840 & -1 &     4.26$\pm$0.55 &     80.7$^{+     27.1} _{-    20.4}$ &    24.38$\pm$0.11 &    27.08$\pm$0.17$^{ V}$ &    26.25$\pm$0.15 &    26.53$\pm$0.38 &    26.03$\pm$0.08$^{V}$  \\ 
N470\_065 &   53.1449 &  -27.7999 &  2.866 & 3 &     5.12$\pm$0.48 &    109.4$^{+     32.4} _{-    25.4}$ &    24.21$\pm$0.09 &    27.82$\pm$0.25$^{ V}$ &    26.28$\pm$0.13 &    26.31$\pm$0.26 &    26.77$\pm$0.18$^{V}$  \\ 
N470\_072 &   53.0783 &  -27.7870 &  2.861 & 2 &     9.82$\pm$0.65 &     99.1$^{+     17.5} _{-    15.7}$ &    23.49$\pm$0.06 &    26.79$\pm$0.12$^{ V}$ &    25.51$\pm$0.09 &    25.21$\pm$0.14 &    25.05$\pm$0.04$^{V}$  \\ 
N470\_076 &   53.1952 &  -27.7793 &  2.848 & 3 &    14.35$\pm$1.18 &     90.6$^{+     18.1} _{-    15.6}$ &    23.07$\pm$0.07 &    25.67$\pm$0.09$^{ V}$ &    25.03$\pm$0.09 &    24.15$\pm$0.08 &    24.19$\pm$0.05$^{V}$  \\ 
N470\_085 &   53.0870 &  -27.7653 &  2.840 & -1 &     7.93$\pm$0.76 &    295.3$^{+    376.2} _{-   117.5}$ &    23.80$\pm$0.09 &    27.23$\pm$0.21$^{ V}$ &    26.42$\pm$0.23 &    26.39$\pm$0.46 &    26.10$\pm$0.10$^{V}$  \\ 
N470\_089 &   53.0968 &  -27.7486 &  2.865 & -1 &     6.86$\pm$0.79 &     54.7$^{+     11.9} _{-    10.3}$ &    23.80$\pm$0.10 &    26.25$\pm$0.09$^{ V}$ &    25.41$\pm$0.10 &    24.95$\pm$0.14 &    25.34$\pm$0.06$^{V}$  \\ 
N470\_092 &   53.1716 &  -27.7434 &  2.856 & -1 &     3.29$\pm$0.64 &     43.1$^{+     13.7} _{-    11.8}$ &    24.54$\pm$0.16 &    26.30$\pm$0.08$^{ V}$ &    25.98$\pm$0.12 &    24.72$\pm$0.08 &    24.06$\pm$0.02$^{V}$  \\ 
N470\_100 &   53.0947 &  -27.7295 &  2.851 & -1 &     3.50$\pm$0.37 &    154.3$^{+     79.1} _{-    46.8}$ &    24.66$\pm$0.10 &    27.29$\pm$0.11$^{ V}$ &    26.94$\pm$0.18 &    26.68$\pm$0.30 &    26.44$\pm$0.09$^{V}$  \\ 
N470\_101 &   53.3347 &  -27.7291 &  2.863 & 3 &     7.91$\pm$0.60 &    103.9$^{+     21.9} _{-    18.9}$ &    23.73$\pm$0.07 &    27.10$\pm$0.34$^{ V}$ &    25.78$\pm$0.10 &    25.78$\pm$0.21 &    25.70$\pm$0.10$^{V}$  \\ 
N470\_111 &   53.0564 &  -27.7094 &  2.846 & 3 &     4.29$\pm$0.38 &   1162.8$^{+   7837.2} _{-   790.2}$ &    24.50$\pm$0.09 &    28.01$\pm$0.25$^{ V}$ &    27.53$\pm$0.31 &    26.47$\pm$0.25 &    26.75$\pm$0.13$^{V}$  \\ 
N470\_116 &   53.0472 &  -27.7042 &  2.842 & 3 &    17.13$\pm$0.71 &    139.2$^{+     19.3} _{-    17.9}$ &    22.92$\pm$0.04 &    26.29$\pm$0.10$^{ V}$ &    25.15$\pm$0.06 &    25.29$\pm$0.15 &    24.94$\pm$0.04$^{V}$  \\ 
N475\_031 &   53.0384 &  -27.9142 &  2.917 & P &     4.05$\pm$0.59 &     28.4$^{+      5.8} _{-     5.5}$ &    24.16$\pm$0.10 &    25.90$\pm$0.07$^{ V}$ &    25.37$\pm$0.07 &    25.04$\pm$0.11 &    25.03$\pm$0.06$^{V}$  \\ 
N475\_069 &   53.0864 &  -27.7814 &  2.886 & -1 &     2.75$\pm$0.49 &     36.1$^{+     10.3} _{-     9.0}$ &    24.65$\pm$0.13 &    26.98$\pm$0.12$^{ V}$ &    26.01$\pm$0.11 &    25.66$\pm$0.17 &    25.48$\pm$0.04$^{V}$ \\ 
N475\_087 &   53.3195 &  -27.7359 &  2.893 & 3 &     3.75$\pm$0.46 &    105.9$^{+     44.9} _{-    30.7}$ &    24.52$\pm$0.11 &    27.20$\pm$0.30$^{ V}$ &    26.59$\pm$0.17 &    27.10$\pm$0.57 &    26.44$\pm$0.16$^{V}$  \\ \hline
                 \multicolumn{11}{c}{ELGs confirmed at $z=2.8$ but not pass the $U-B$ cut.} \\\hline
N470\_060 &   53.1478 &  -27.8128 &  2.851 & -1 &     3.31$\pm$0.40 &    143.8$^{+     80.5} _{-    46.5}$ &    24.71$\pm$0.11 &    27.16$\pm$0.12$^{ V}$ &    26.95$\pm$0.20 &    27.87$\pm$0.92 &    26.78$\pm$0.18$^{V}$  \\ 
N470\_068 &   53.2288 &  -27.7985 &  2.856 & 3 &     3.96$\pm$0.47 &    130.6$^{+     56.4} _{-    37.9}$ &    24.51$\pm$0.11 &    26.96$\pm$0.12$^{ V}$ &    26.69$\pm$0.17 &    26.08$\pm$0.20 &    25.88$\pm$0.07$^{V}$  \\ 
N470\_077 &   52.9808 &  -27.7786 &  2.848 & 3 &     3.78$\pm$0.45 &     89.3$^{+     28.4} _{-    21.6}$ &    24.52$\pm$0.11 &    26.49$\pm$0.07$^{ V}$ &    26.46$\pm$0.14 &    26.22$\pm$0.23 &    25.91$\pm$0.10$^{V}$  \\ 
N475\_110 &   53.3013 &  -27.6872 &  2.924 & 2 &     5.02$\pm$0.65 &     76.8$^{+     27.4} _{-    20.7}$ &    24.16$\pm$0.11 &    25.71$\pm$0.11$^{ V}$ &    26.03$\pm$0.16 &    25.68$\pm$0.23 &    25.77$\pm$0.11$^{V}$  \\ 
N466\_074 &   53.0438 &  -27.8625 &  2.837 & 2 &     5.99$\pm$0.85 &     45.0$^{+     10.6} _{-     9.5}$ &    23.90$\pm$0.11 &    25.61$\pm$0.07$^{ V}$ &    25.37$\pm$0.09 &    25.14$\pm$0.15 &    25.04$\pm$0.07$^{V}$  \\ 
N475\_042 &   53.0628 &  -27.8627 &  2.907 & 2 &     4.16$\pm$0.49 &     33.8$^{+      6.2} _{-     5.9}$ &    24.19$\pm$0.08 &    25.35$\pm$0.04$^{ V}$ &    25.51$\pm$0.07 &    24.95$\pm$0.09 &    24.83$\pm$0.04$^{V}$  \\ 
N466\_073 &   53.2261 &  -27.8637 &  2.818 & 3 &     5.20$\pm$0.53 &     51.1$^{+      8.5} _{-     8.0}$ &    24.08$\pm$0.08 &    25.69$\pm$0.03$^{ V}$ &    25.65$\pm$0.07 &    25.19$\pm$0.09 &    25.35$\pm$0.04$^{V}$  
\enddata
  \tablenotetext{a}{`Q' presents the spectral quality. Here 'Q = 	P' means spectral information from public surveys, and negative Q value means that the spectrum is affected by slit overlap or CCD gap. Q = 2 means that the emission line feature is weak but visible. Q $\geq$ 3 means that the emission line feature is strong and significant. The \lya\ line of quasar N466\_109 is the brightest \lya\ line and we mark it with Q=4.    } 
\tablenotetext{b}{ The \lya\ line flux is calculated from the narrow-band and broad-band photometry in the unit of 10$^{-17}$ erg s$^{-1}$ cm$^{-2}$.  }
\tablenotetext{c}{The $U$ or $R$ band magnitudes of LAEs covered by VIMOS-U or VIMOS-R band are updated.  Here superscript '$V$' or '$M$' present photometric values from VIMOS or MUSYC data.  }
\tablenotetext{d}{Obj. N466\_109 is the BAL-QSO at z=2.81.  }
\end{deluxetable*}

\begin{deluxetable*}{lccccccccl}
\tabletypesize{\scriptsize}
\tablecaption{The {\it GALEX} UV and {\it Chandra} X-ray  detections of emission line galaxies selected in our narrowband images.\label{xdet}}
\tablewidth{0pt}
\tablehead{
\colhead{Obj.} & \colhead{RA} & \colhead{DEC} & \colhead{$U-B$} & \colhead{$z$} & \colhead{F $\times$10$^{-17}$} & \colhead{EW$_{obs}$(line) } & \colhead{Match-ID} &
\colhead{Sep.} & \colhead{Note} \\
                 &                &            &         &         &   [\,erg\,cm$^{-2}$\,s$^{-1}$]      & [\AA]        &               &  [\arcsec]  &   
}
\startdata
                                  \multicolumn{10}{c}{\it{GALEX} UV detections in ECDFS} \\ \hline
N466\_007 &   53.3838  &  -28.0303 &  0.09 & --  &     5.52$\pm$0.93 &    181.8$^{+     54.2} _{-    71.9}$ & GALEXJ033332.0-280148 &     0.25 &  \\ 
N466\_015 &   53.2585  &  -27.9991 & -0.36 & --  &     2.81$\pm$0.68 &     81.5$^{+     23.2} _{-    25.9}$ & GALEXJ033302.0-275956 &     0.30 &  \\ 
N466\_037 &   53.0707  &  -27.9235 &  0.06 & --  &     6.84$\pm$0.60 &     84.3$^{+      8.8} _{-     9.1}$ & GALEXJ033217.0-275524 &     0.41 &  \\ 
N466\_061 &   52.9714  &  -27.8839 & -0.03 & --  &     2.66$\pm$0.69 &     92.3$^{+     27.4} _{-    31.3}$ & GALEXJ033153.0-275301 &     0.79 &  \\ 
N466\_062 &   53.0305  &  -27.8777 &  0.10 & --  &     2.87$\pm$0.58 &    111.2$^{+     27.5} _{-    30.3}$ & GALEXJ033207.3-275239 &     0.33 &  \\ 
N466\_124 &   53.1446  &  -27.7855 &  0.18 & --  &    20.96$\pm$0.68 &     82.6$^{+      3.2} _{-     3.2}$ & GALEXJ033234.7-274707 &     0.17 &  \\ 
N466\_152 &   53.3062  &  -27.7455 & -0.27 & --  &     1.94$\pm$0.62 &     78.4$^{+     27.5} _{-    31.9}$ & GALEXJ033313.4-274444 &     0.58 &  \\ 
N470\_009 &   53.2495  &  -27.9900 & -0.22 & --  &     4.31$\pm$1.00 &     82.0$^{+     22.5} _{-    24.7}$ & GALEXJ033259.8-275924 &     0.78 &  \\ 
N470\_052 &   53.0412  &  -27.8310 & -0.29 & --  &     2.84$\pm$0.58 &     87.2$^{+     21.5} _{-    23.6}$ & GALEXJ033209.8-274951 &     0.46 &  \\ 
N470\_086 &   53.3636  &  -27.7577 &  0.82 & --  &     2.36$\pm$0.55 &    133.8$^{+     40.1} _{-    47.5}$ & GALEXJ033327.2-274528 &     0.44 &  \\ 
N470\_115$^{*}$ &   53.2526  &  -27.7042 &  0.59 & 0.26  &     5.48$\pm$0.96 &    106.4$^{+     23.0} _{-    25.0}$ & GALEXJ033300.6-274214 &     0.49 & [O{\sc ii}] emitter  \\ 
N470\_119 &   53.0811  &  -27.6980 & -0.40 & --  &     3.91$\pm$1.11 &     87.1$^{+     28.1} _{-    33.1}$ & GALEXJ033219.4-274152 &     0.86 &  \\ 
N475\_027 &   53.0514  &  -27.9327 & -0.07 & --  &     2.50$\pm$0.61 &     95.9$^{+     28.5} _{-    32.5}$ & GALEXJ033212.3-275557 &     0.53 &  \\ 
N475\_086 &   53.2092  &  -27.7400 & -0.52 & 0.5616 &     2.51$\pm$0.65 &    174.7$^{+     63.1} _{-    82.6}$ & GALEXJ033250.1-274424 &     0.81 &  \\ 
N475\_117 &   53.2545  &  -27.6722 & -0.07 & --  &     2.55$\pm$0.49 &    199.0$^{+     56.9} _{-    71.0}$ & GALEXJ033301.1-274020 &     0.62 &  \\ 
N475\_123 &   52.8566  &  -27.6600 & -0.82 & --  &     5.91$\pm$0.92 &    343.9$^{+    155.8} _{-   616.6}$ & GALEXJ033125.5-273936 &     0.34 &  \\ 
N475\_057$^{*}$ &   53.0803  &  -27.8157 &  0.30 & 0.677  &    15.12$\pm$0.99 &     89.4$^{+      7.9} _{-     7.6}$ & GALEXJ033219.2-274857 &     0.71 & Mg{\sc ii} \& X-det. \\ 
N475\_128 &   53.0169  &  -27.6238 & -1.13 &0.977 &    24.38$\pm$0.58 &    233.3$^{+     11.5} _{- 10.9}$ & GALEXJ033204.0-273725 &     0.19 &  all NBs \& X-det. \\ \hline
                 \multicolumn{10}{c}{{\it Chandra} X-ray  detections in CDFS (Xue et al. 2011)} \\\hline
N466\_038 &   53.2531  &  -27.9224 & -0.27 & 2.005  &    49.57$\pm$0.81 &     87.2$^{+      1.7} _{-     1.7}$ & CDFS\_691 &     0.85 & C{\sc iv} in NB466  \\ 
N475\_064 &   52.9681  &  -27.7980 &  0.27 & --  &     2.05$\pm$0.53 &    100.9$^{+     31.9} _{-    37.5}$ & CDFS\_39 &     0.35 &  \\ 
N466\_109$^{**}$ &   53.0393  &  -27.8019 &  1.86 & 2.81  &   171.29$\pm$0.94 &    117.7$^{+      0.9} _{-     0.8}$ & CDFS\_149 &     0.63 & BAL Quasar  \\ 
N475\_057 &   53.0803  &  -27.8157 &  0.30 & 0.677 &    15.12$\pm$0.99 &     89.4$^{+      7.9} _{-     7.6}$ & CDFS\_270 &     0.54 &  Mg{\sc ii}  \& GUV-det. \\ 
N475\_128 &   53.0169  &  -27.6238 & -1.13 & 0.977  &    24.38$\pm$0.58 &    233.3$^{+     11.5} _{-    10.9}$ & CDFS\_101 &     0.45 &  all NBs \& GUV-det \\ \hline
                 \multicolumn{10}{c}{ {\it Chandra} X-ray  detections in ECDF-S (Lehmer et al. 2006)} \\\hline
N466\_026$^{**}$ &   53.3118  &  -27.9636 &  0.65 & --  &    26.30$\pm$0.71 &    206.7$^{+      9.3} _{-     9.1}$ & ECDFS\_625 &     0.47 &  \\ 
N466\_141 &   52.8912  &  -27.7672 &  1.87 & --  &     1.79$\pm$0.51 &     85.0$^{+     27.2} _{-    31.4}$ & ECDFS\_89 &     0.22 &  \\ 
N466\_211$^{**}$ &   52.9954  &  -27.5878 &  0.92 & --  &    54.49$\pm$0.78 &    174.5$^{+      4.2} _{-     3.9}$ & ECDFS\_246 &     0.39 &  \\ 
N475\_029 &   52.9079  &  -27.9263 &  0.09 & --  &     3.13$\pm$0.72 &    111.6$^{+     32.8} _{-    37.7}$ & ECDFS\_109 &     0.54 &  \\ 
N475\_131 &   52.9030  &  -27.5793 &  0.18 & --  &     7.93$\pm$0.51 &    103.2$^{+      9.4} _{-     9.2}$ & ECDFS\_102 &     0.20 &  
\enddata
\tablenotetext{$*$}{The two LAE candidates are excluded from the complete LAE sample because of their GUV detections.}
\tablenotetext{$**$}{The three X-ray detected LAE candidates are kept in the complete LAE sample.}
\end{deluxetable*}

\begin{deluxetable*}{lccccccccl}
\tabletypesize{\scriptsize}
\tablecaption{The match of our emission line candidates with public spectroscopic surveys. \label{ecdet}}
\tablewidth{0pt}
\tablehead{
\colhead{Obj.} & \colhead{RA} & \colhead{DEC} & \colhead{$U-B$} & \colhead{$z$(public)} & \colhead{F(Line)} & \colhead{EW$_{obs}$(line) } & \colhead{Match-ID [Ref]$^+$}  &
\colhead{Sep.} & \colhead{Note} \\
                 &                &            &         &    --$_{[quality]}$     &   [10$^{-17}$]      & [\AA]        &               &  [\arcsec]  &   
}
\startdata
                 \multicolumn{10}{c}{{\it LAE candidates in the complete sample with secure redshifts at $z\sim$ 2.8--2.9}} \\\hline
   N466\_210   & 53.01274 & -27.60039   & 0.36   & 2.81$_{[3]}$ &   9.75$\pm$1.35  &        171.3$^{+40.1}_{-35.5}$             &     -- [20]&   0.62  & \\ 
   N466\_069      & 53.02648 & -27.86803   & 0.41   & 2.805$_{[3]}$ &    13.5$\pm$0.9   &  351.4$^{+56.7}_{-51.4}$               &     -- [20] &   0.14   &  $z_{\textsc{IM}}$ = 2.814 \\ 
  N475\_031 & 53.03839 & -27.91416 & 0.53 & 2.917$_{ [A]}$ &4.05$\pm$0.59 & 111.0$^{+22.9}_{-21.6}$ & {\tiny GOODS\_LRb\_001\_1\_q3\_67\_2} [14] & 0.24 &  \\
N466\_109 & 53.03933 & -27.80191 & 1.86 & 2.817$_{ \textsc{[--]}}$ &171.29$\pm$0.94 & 117.7$^{+0.8}_{-0.9}$ & 5 [3] & 0.13 &$z_{\textsc{IM}}$ = 2.817   \\
N466\_072 & 53.04612 & -27.86374 & 0.51 & 2.805$_{ [B]}$ &2.89$\pm$0.40 & 163.0$^{+36.2}_{-33.1}$ & {\tiny GOODS\_LRb\_001\_1\_q3\_90\_1} [14] & 0.76 & $z_{\textsc{IM}}$ = 2.817  \\
N466\_105 & 53.05163 & -27.80457 & 0.84 & 2.812$_{ [A]}$ &17.30$\pm$0.87 & 187.0$^{+14.8}_{-15.1}$ & {\tiny GOODS\_LRb\_001\_q2\_9\_1} [14] & 0.22 & $z_{\textsc{IM}}$ = 2.818  \\
 N466\_086     & 53.05735 & -27.84084   & 0.61   & 2.8256$_{[3]}$ &   4.1$\pm$0.8      &  111.9$^{+29.9}_{-27.0}$                   &     -- [20] &   0.45   &  $z_{\textsc{IM}}$ = noLine \\                  
 N466\_096 & 53.08630 & -27.81870 & 0.32 & 2.813$_{ [A]}$ &7.13$\pm$0.91 & 160.9$^{+32.2}_{-30.0}$ & {\tiny GOODS\_LRb\_dec06\_2\_q3\_51\_1} [14] & 0.66 &  \\
N475\_045 & 53.19309 & -27.84808 & 1.20 & 2.898$_{ [A]}$ &4.55$\pm$0.64 & 73.6$^{+12.8}_{-12.4}$ & {\tiny GOODS\_LRb\_001\_1\_q4\_64\_1} [14] & 0.07 &  \\
N466\_178 & 53.20657 & -27.70322 & 0.64 & 2.821$_{ [A]}$ &17.36$\pm$0.99 & 271.3$^{+30.0}_{-28.5}$ & {\tiny GOODS\_LRb\_001\_q1\_35\_1} [14] & 0.40 &$z_{\textsc{IM}}$ = 2.823   \\
N466\_164 & 53.21243 & -27.72730 & 0.71 & 0.404$^*_{ [A]}$ &5.07$\pm$0.58 & 151.8$^{+24.4}_{-23.4}$ & {\tiny GOODS\_LRb\_001\_q1\_32\_1} [14] & 0.25 & $z_{\textsc{IM}}$ = 2.822  \\
N466\_187 & 52.99377 & -27.68139 & 0.98 & 2.801$_{ [C]}$ &5.12$\pm$0.92 & 92.2$^{+21.8}_{-20.2}$ & {\tiny GOODS\_LRb\_003\_new\_2\_q2\_28\_1} [14] & 0.51 &  \\\hline
                 \multicolumn{10}{c}{{\it LAE candidates in the complete sample with other redshifts}} \\\hline                
 N470\_013 & 53.02760 & -27.97011 & 0.64 & 1.091$_{ [3]}$ &3.47$\pm$0.70 & 103.6$^{+28.1}_{-25.6}$ & 10862 [1] & 0.44 &  \\
 N470\_117 & 53.06053 & -27.69862 & 3.07 & 0.416$_{ [B]}$ &3.01$\pm$0.55 & 177.7$^{+55.4}_{-47.5}$ & {\tiny GOODS\_LRb\_001\_q2\_45\_1} [14] & 0.36 & $z_{\textsc{IM}}$ = lowzE  \\
N475\_057 & 53.08031 & -27.81572 & 0.30 & 0.677$_{ [B]}$ &15.12$\pm$0.99 & 89.4$^{+7.6}_{-7.9}$ & {\tiny GOODS\_MR\_dec06\_3\_q2\_34\_1} [14] & 0.32 & GUV  \\
N470\_093 & 53.21309 & -27.74215 & 5.62 & 0.534$_{ [4]}$ &4.36$\pm$0.78 & 128.1$^{+32.5}_{-29.2}$ & 75367  [1] & 0.23 &  \\
N470\_115 & 53.25259 & -27.70418 & 0.59 & 0.265$_{ [A]}$ &5.48$\pm$0.96 & 106.4$^{+25.1}_{-23.0}$ & {\tiny GOODS\_LRb\_dec06\_3\_q1\_69\_1} [14] & 0.51 &  GUV, $z_{\textsc{IM}}$ = lowzE\\
N466\_158 & 52.94563 & -27.73616 & 0.77 & 2.737$_{ [C]}$ &3.58$\pm$0.85 & 114.1$^{+38.3}_{-33.2}$ & {\tiny GOODS\_LRb\_001\_1\_q2\_38\_1} [14] & 0.67 &  \\
N475\_016 & 53.13192 & -27.97158 & 0.94 & 2.343$_{ [C]}$ &3.41$\pm$0.62 & 122.7$^{+34.2}_{-30.6}$ & {\tiny GOODS\_LRb\_001\_q3\_26\_1} [14] & 0.66 &  \\\hline
                 \multicolumn{10}{c}{{\it LAE candidates with F$(Ly\alpha)$ $\leq$ 2.9$\times$10$^{10^{-17} }$ erg s$^{-1}$ cm$^{-2}$}} \\\hline                
N470\_042 & 53.24130 & -27.88101 & 0.44 & 2.339$_{ [B]}$ &2.33$\pm$0.53 & 92.6$^{+27.9}_{-25.3}$ & {\tiny GOODS\_LRb\_001\_q4\_8\_1} [14] & 0.10 & $z_{\textsc{IM}}$ = lowzC \\
N475\_066 & 53.25439 & -27.79280 & 1.67 & 3.334$_{ [B]}$ &1.94$\pm$0.53 & 84.5$^{+30.7}_{-26.9}$ & {\tiny GOODS\_LRb\_dec06\_2\_q4\_53\_1} [14] & 0.16 &  \\
N475\_035 & 53.06929 & -27.89995 & 0.37 & 2.566$_{ [A]}$ &2.72$\pm$0.55 & 103.8$^{+30.0}_{-27.1}$ & {\tiny GOODS\_LRb\_001\_1\_q3\_95\_3} [14] & 0.37 &  \\               
N475\_046 & 53.00151 & -27.84772 & 0.66 & 2.905$_{ [C]}$ &2.04$\pm$0.40 & 74.7$^{+18.6}_{-17.5}$ & {\tiny GOODS\_LRb\_003\_new\_2\_q3\_3\_1} [14] & 0.14 &  \\
N475\_081 & 53.15006 & -27.75226 & 1.96 & 0.000$_{ [C]}$ &2.22$\pm$0.34 & 203.7$^{+62.2}_{-49.5}$ & {\tiny GOODS\_LRb\_002\_1\_q1\_49\_1} [14] & 0.30 &  \\\hline
                 \multicolumn{10}{c}{{\it ELGs NOT passed LAE selection  }} \\\hline
N466\_046 & 52.96759 & -27.90259 & 0.14 & 0.246$_{ [4]}$ &4.59$\pm$0.50 & 86.6$^{+11.7}_{-11.4}$ & 16549 [1] & 0.42 &  \\
N475\_128 & 53.01687 & -27.62375 & -1.13 & 0.977$_{ [2]}$ &24.38$\pm$0.58 & 233.3$^{+11.0}_{-11.5}$ & 539 [2] & 0.03 & GUV \& X-ray  \\
N466\_180 & 53.02215 & -27.70129 & -0.07 & 0.861$_{ [3]}$ &3.02$\pm$0.62 & 119.6$^{+34.4}_{-30.5}$ & 34011 [1] & 0.66 & $z_{\textsc{IM}}$ = lowzC   \\
N466\_037 & 53.07072 & -27.92353 & 0.06 & 0.250$_{ [A]}$ &6.84$\pm$0.60 & 84.3$^{+9.1}_{-8.8}$ & {\tiny GOODS\_LRb\_003\_new\_2\_q3\_20\_1} [14] & 0.26 & GUV \\
N470\_097 & 53.15581 & -27.73532 & -0.25 & 1.017$_{ [1]}$ &3.87$\pm$0.81 & 80.2$^{+21.1}_{-19.2}$ & CDFS\_00344 [15] & 0.27 &  \\
N466\_150 & 53.20231 & -27.75137 & 0.24 & 1.113$_{ [2]}$ &1.95$\pm$0.54 & 84.3$^{+29.1}_{-26.1}$ & 29376 [1] & 0.18 &  \\
N475\_082 & 53.20435 & -27.74988 & 0.28 & 0.156$_{ [A]}$ &3.05$\pm$0.67 & 90.3$^{+26.1}_{-23.5}$ & {\tiny GOODS\_LRb\_001\_q1\_19\_1} [14] & 0.43 &  \\
N475\_086    & 53.2092 & -27.7400   & -0.52   & 0.5616$_{[3]}$ &    2.51$\pm$0.65  &   174.7$^{+82.6}_{-63.1}$       &     -- [20] &   0.52   &  GUV, $z_{\textsc{IM}}$ = lowzE \\ 
N466\_038 & 53.25307 & -27.92239 & -0.27 & 2.005$_{ [A]}$ &49.57$\pm$0.81 & 87.2$^{+1.7}_{-1.7}$ & {\tiny GOODS\_LRb\_003\_new\_q4\_1\_1} [14] & 0.42 & X-ray  \\
N466\_104 & 53.26567 & -27.80572 & 0.28 & 2.808$_{ [B]}$ &6.24$\pm$0.76 & 77.9$^{+11.9}_{-11.4}$ & {\tiny GOODS\_LRb\_003\_new\_q4\_32\_1} [14] & 0.62 &  
\enddata
\tablenotetext{$^*$}{We checked our spectrum and the public spectrum of N466\_164. We judge that N466\_164 is a LAE at $z$ = 2.822.   }
\tablenotetext{$^+$}{The references here are same as the CDFS master catalog. 1: Le Fevre et al. (2005); 2: \citet{Szokoly04}; 3: Croom et al. 2001; 14: \citet{Popesso09,Balestra10}; 15: Mignoli et al. (2005); 20: \citep{LeFevre15}    }
\end{deluxetable*}

\begin{deluxetable*}{lcccccc}
\tabletypesize{\scriptsize}
\tablecaption{Summary of the \lya\ luminosity functions at $z\sim$2.8  \label{lyalfs}}
\tablewidth{0pt}
\tablehead{Field & \colhead{ $z$ } & \colhead{log$_{10}$($\Phi^*$) } & \colhead{log$_{10}$($L^*_{Ly\alpha}$)}  & \colhead{$\alpha$} & \colhead{$\chi^2$} & \colhead{log$_{10}$($\Psi_{Ly\alpha}$)}  \\
}
\startdata
All NBs              &  2.86$\pm$0.06          &  -3.21$\pm$0.11                                 &    42.73$\pm$0.08                                & -1.5 (fix)                             & 18.5/10                   &    -2.37$\pm$0.10  \\
NB466           &    2.82$\pm$0.02           &  -3.22$\pm$0.15                              & 42.84$\pm$0.13                               & -1.5 (fix)                            &    11.0/8                         &        -2.27$\pm$0.14  \\
NB470+NB475 & 2.88$\pm$0.04         & -3.04$\pm$0.17                                & 42.49$\pm$0.10                              & -1.5 (fix)                            &   12.6/9                         &       -2.44$\pm$0.15  \\
\enddata
\end{deluxetable*}






\end{document}